\documentclass[12pt]{article}
\usepackage{geometry}
\usepackage{amsmath}
\usepackage{amssymb,epsfig,subfigure}
\usepackage{graphicx}
\numberwithin{equation}{section}

\newcommand{\be}{\begin{equation}}
\newcommand{\ee}{\end{equation}}
\newcommand{\ba}{\begin{aligned}}
\newcommand{\ea}{\end{aligned}}
\newcommand{\half}{{1\over 2}}

\def\m1{\left(-1\right)^{F_i}}

\makeatletter
\def\sla@#1#2#3#4#5{{%
  \setbox\z@\hbox{$\m@th#4#5$}%
  \setbox\tw@\hbox{$\m@th#4#1$}%
  \dimen4\wd\ifdim\wd\z@<\wd\tw@\tw@\else\z@\fi
  \dimen@\ht\tw@
  \advance\dimen@-\dp\tw@
  \advance\dimen@-\ht\z@
  \advance\dimen@\dp\z@
  \divide\dimen@\tw@
  \advance\dimen@-#3\ht\tw@
  \advance\dimen@-#3\dp\tw@
  \dimen@ii#2\wd\z@  \raise-\dimen@\hbox to\dimen4{%
    \hss\kern\dimen@ii\box\tw@\kern-\dimen@ii\hss}%
  \llap{\hbox to\dimen4{\hss\box\z@\hss}}}}
\def\slashed#1{%
  \expandafter\ifx\csname sla@\string#1\endcsname\relax
    {\mathpalette{\sla@/00}{#1}}%
  \else
    \csname sla@\string#1\endcsname
  \fi}
\makeatother

\begin{document}


\thispagestyle{empty}
\begin{flushright}\footnotesize
\texttt{CALT-68-2726}\\
\vspace{2.1cm}
\end{flushright}

\renewcommand{\thefootnote}{\fnsymbol{footnote}}
\setcounter{footnote}{0}

\begin{center}
{\Large\textbf{\mathversion{bold}F-theory Compactifications for Supersymmetric GUTs}\par}

\vspace{2.1cm}

\textrm{Joseph Marsano, Natalia Saulina and Sakura Sch\"afer-Nameki}

\vspace{1cm}

\textit{California Institute of Technology\\
1200 E California Blvd., Pasadena, CA 91125, USA } \\
\texttt{marsano, saulina, ss299  theory.caltech.edu}

\bigskip


\par\vspace{1cm}

\textbf{Abstract}\vspace{5mm}
\end{center}

\noindent
We construct a family of elliptically fibered Calabi-Yau four-folds $Y_4$ for F-theory compactifications that realize $SU(5)$ GUTs in the low-energy limit.  The three-fold base $X_3$ of these fibrations is almost Fano and satisfies the topological criteria required to ensure that the $U(1)_Y$ gauge boson remains massless, while allowing a decoupling of GUT and Planck scale physics.  We study generic features of these models and the ability to engineer three chiral generations of MSSM matter.  Finally, we demonstrate that it is relatively easy to implement the topological conditions required to reproduce certain successful features of local F-theory models, such as the emergence of flavor hierarchies.

\vspace*{\fill}

\setcounter{page}{1}
\renewcommand{\thefootnote}{\arabic{footnote}}
\setcounter{footnote}{0}

 \newpage


\tableofcontents

\newpage

\section{Introduction}


The past year has seen incredible progress in the study of F-theory realizations of supersymmetric GUTs
\cite{Donagi:2008ca,Beasley:2008dc, Beasley:2008kw,Donagi:2008kj}.  Much of this work has proceeded in the general spirit of bottom-up phenomenology, focusing on the study of local models based on noncompact Calabi-Yau four-folds.  This has enabled a number of phenomenological issues to be addressed, including supersymmetry-breaking \cite{Heckman:2008es,Marsano:2008py} and gauge mediation \cite{Marsano:2008jq,Heckman:2008qt}, as well as flavor structure \cite{Font:2008id,Heckman:2008qa,Hayashi:2009ge,Heckman:2009de} and models of neutrino physics \cite{Bouchard:2009bu,Randall:2009dw}.  Quite nicely, the models that emerge most naturally in local F-theory constructions, such as \cite{Marsano:2008jq} and \cite{Heckman:2008qt}, give rise to 
effective field theories which exhibit several characteristic features, including a large messenger scale and a relatively heavy gravitino \cite{Ibe:2007km, Heckman:2008qt}.  Studies of the relevant gravitino cosmology include \cite{Ibe:2006rc} and \cite{Heckman:2008jy}, while collider signatures of these models have been investigated in \cite{Ibe:2007km} and \cite{Heckman:2009bi}. Other recent work on F-theory model building includes \cite{Braun:2008pz, Blumenhagen:2008aw,  Bourjaily:2009vf, Laamara:2009gd, Chen:2009me}.

 
One expects, however, that combining local phenomenological requirements with the global consistency conditions needed to embed them into full F-theory compactifications provides a highly constrained setup.  While local models help to identify specific geometric structures that are desirable for model building, actually realizing such structures, if possible at all, will likely require significant tuning.  Recently, some general properties of F-theory compactifications that give rise to $SU(5)$ GUT models have been described in \cite{Andreas:2009uf}.  Further, a rigorous approach to building local models that incorporates many of the constraints needed to compactify them has recently been developed in \cite{Donagi:2009ra}.  The techniques described therein rely heavily on input from heterotic models, though the work of \cite{Hayashi:2009ge,Donagi:2009ra} demonstrates that they can be applied more generally to F-theory compactifications that do not admit heterotic duals.  This is important because the most successful mechanism for breaking the GUT gauge group is absent in models with a dual heterotic description.  Some aspects of the heterotic/F-theory duality map that helped elucidate important structures in F-theory have been discussed recently in \cite{Hayashi:2008ba, Hayashi:2009ge}.

Despite the ability to engineer GUTs based on $SO(10)$ or $E_6$ gauge groups in F-theory, problems with charged exotics have led to a particular emphasis on $SU(5)$ models \cite{Beasley:2008kw,Donagi:2008kj}.  Theories of this type can be engineered directly in the orientifold limit of F-theory.  Indeed, compact type IIB orientifold models for $SU(5)$ GUTs that implement the same GUT-breaking mechanism as their F-theory counterparts have been constructed in the past year \cite{Blumenhagen:2008zz}.  What motivates us to consider F-theory compactifications, however, is rather the favorable flavor structure that emerges \cite{Heckman:2008qa}.  Type IIB models such as those of \cite{Blumenhagen:2008zz} require nonperturbative effects to generate the up-type Yukawa couplings, suggesting that an intrinsically nonperturbative framework may be more appropriate.  F-theory models, on the other hand, treat up- and down-type Yukawa couplings on an equal footing and can lead to natural hierarchies for both \cite{Heckman:2008qa}.




In this paper, we construct F-theory compactifications that engineer supersymmetric GUTs and work towards the further realization of as many of the desirable phenomenological features of local models as possible.  The most important properties that should be realized are an $SU(5)$ gauge group along with a mechanism for breaking this group down to the MSSM gauge group  $SU(3)\times SU(2)\times U(1)_Y$.  The $SU(5)$ gauge bosons are localized on a four-cycle, $S_{\rm GUT}$, of the compact geometry that corresponds to an $I_5=SU(5)$ singular locus of the elliptic fibration{\footnote{Throughout this paper, we will use ADE groups to label such singularities rather than the Kodaira notation.}}.  A simple mechanism for breaking the GUT gauge group,  is the introduction of a nontrivial hypercharge flux along $S_{\rm GUT}$ \cite{Beasley:2008kw,Donagi:2008kj}.  If $S_{\rm GUT}$ is a del Pezzo surface, the bundle cohomologies that determine the zero mode structure exhibit a number of vanishing theorems that facilitate the lifting of charged exotics \cite{Beasley:2008kw}.  An internal flux will generically give a mass to the hypercharge gauge boson, though, unless an important topological constraint, that we discuss in detail, is satisfied \cite{Buican:2006sn,Beasley:2008kw,Donagi:2008kj}.  

Our first goal in the current work is the construction of a simple three-fold, $X_3$, that can serve as the base of elliptically fibered Calabi-Yau four-folds for F-theory compactifications that realize this mechanism of GUT-breaking. Furthermore, there should exist a limit in which GUT and Planck scale physics decouple, which is ensured if the surface $S_{\rm GUT}$ is contractible \cite{Beasley:2008kw}.
 In particular, we construct a three-fold that exhibits the following three properties\footnote{In section
4.4 we construct an example of $X_3$ with non-contractible $S_{GUT}=dP_2.$ However, we show that the volume of $X_3$
can be much greater than the volume of $S_{GUT}$ which allows for mass hierarchy $M_{GUT}\ll M_{pl}$.} 

\begin{itemize}
\item $X_3$ is almost Fano, i.e. $K_{X_3}^{-1}$ is almost ample: $K_{X_3}^{-1} \cdot C \leq 0$ for all curves $C$
\item $X_3$ contains a contractible del Pezzo divisor that serves as $S_{\rm GUT}$
\item With this choice of $S_{\rm GUT}$, $X_3$ satisfies the topological constraint that ensures a massless $U(1)_Y$ gauge boson
\end{itemize}
We focus in this paper on a particular $X_3$ but the method that we use to construct it can be applied more generally than we do here.  The specific construction that we consider is a blow-up of a nodal curve, which appears in Mori's classification of Fano three-folds \cite{Mori} (see also \cite{Matsuki}, as well as \cite{Buican:2006sn} in the context of Calabi-Yau three-folds), followed by a flop transition.  The resulting geometry can be described purely algebraically and does not seem to have a toric realization.

Once we have our three-fold $X_3$, we next turn to the study of generic elliptic fibrations that exhibit an $SU(5)$ singularity along the del Pezzo surface, $S_{\rm GUT}$.  Charged matter in these models is localized on various matter curves inside $S_{\rm GUT}$ whose structure and intersections are easy to determine from the topological data of $X_3$ \cite{Andreas:2009uf}.  The matter curve on which fields transforming in the $\mathbf{10}$ of $SU(5)$ are localized is fairly simple and has the topology of a $\mathbb{P}^1$.  Matter in the $\overline{\mathbf{5}}$, however, localizes on a much more complicated matter curve that is of significantly higher genus in general.  In addition to studying the matter curves, we show that these compactifications satisfy all of the requisite global consistency constraints and demonstrate, using the techniques of \cite{Donagi:2009ra}, that a suitable tuning permits us to turn on enough (globally well-defined) $G$-fluxes to obtain three chiral generations of MSSM matter fields.  We therefore obtain a fairly large family of honest F-theory compactifications that realize three-generation $SU(5)$ GUTs.

Favorable phenomenology, however, requires much more structure.  Local models typically exhibit a very precise and intricate pattern of matter curves and intersections in order to engineer realistic effective field theories.  In this paper, we work towards the realization of two specific successes of local models, namely the extension of the proton lifetime and the realization of natural hierarchies in the up- and down-type Yukawa matrices.  Implementing these features requires two steps.  First, one must impose a number of geometric constraints which amount to requiring the $\mathbf{5}$ matter curve to factorize into a number of components with specific intersection properties.  This will provide us with candidate components on which to engineer the Higgs doublets, $H_u$ and $H_d$, and the $\mathbf{\overline{5}}$ matter fields.  To complete the model, however, one must move to the second step.  That is, one must demonstrate that it is possible to turn on suitable $G$-fluxes that not only give rise to 3 chiral generations but also causes them to localize on the ``correct" components of $\Sigma_5$.  In this paper, we address only the first step and demonstrate that candidate four-folds with the right geometric properties are not hard to explicitly construct.  We will return to the second issue in future work.  Eventually, it will be important to address issues related to moduli stabilization as well.   We do not address this issue in the present work.

The outline of our paper is as follows.  We begin in section \ref{sec:Loc} by reviewing the basic properties of $SU(5)$ GUTs and the properties that four-folds must exhibit in order to reproduce various phenomenological successes of local models.  In section \ref{sec:Global} we discuss important global consistency constraints, including tadpole cancellation and $G$-flux quantization conditions. We then turn to the main constructions of the paper in section \ref{sec:ThreeFold}, where we obtain an almost Fano three-fold $\tilde{X}$ that will serve as the base for our elliptic fibrations. In section \ref{sec:SU5} we study generic elliptic fibrations over $\tilde{X}$ that yield $SU(5)$ GUTs and demonstrate, among other things, that under a suitable tuning it is possible to obtain three-generation models.  We then describe how to implement geometric refinements that incorporate some crucial features of local models in section \ref{sec:Realistic}.  Various details of the geometries are discussed in the appendices.


{\bf Note added:} While this paper was in preparation the preprint \cite{Donagi:2009ra} by R.~Donagi and M.~Wijnholt appeared, which also discusses issues related to compact models in F-theory.  After \cite{Donagi:2009ra} appeared, we applied their techniques for studying $G$-fluxes to obtain the results of subsection \ref{sec:ThreeGen}.





\section{Input from Local Models}
\label{sec:Loc}

Recent studies of local models in F-theory have demonstrated that it is possible to realize a number of phenomenologically desirable features \cite{Beasley:2008kw,Donagi:2008kj,Marsano:2008jq,Heckman:2008qt,Font:2008id,Heckman:2008qa,Bouchard:2009bu,Randall:2009dw}.  This includes not only the presence of a suitable GUT gauge group and MSSM matter content but also a viable method of GUT-breaking \cite{Beasley:2008kw,Donagi:2008kj}, a mechanism for achieving doublet-triplet splitting without violating current constraints from proton decay experiments \cite{Beasley:2008kw}, and natural hierarchies in the Yukawa matrices \cite{Heckman:2008qa, Bouchard:2009bu,Randall:2009dw}.  Because we would like to build compactifications that are as realistic as possible, we review in this section the basic structures of local models in F-theory that give rise to these features.

\subsection{GUT Breaking and Hypercharge Flux}\label{subsec:GUTbreaking}

In F-theory models, the GUT gauge group is realized on a stack of 7-branes that wraps a four-cycle, $S_{\rm GUT}$, in the compactification geometry.  A promising technique for breaking the GUT group is the introduction of internal $U(1)$ fluxes.  Suitably chosen, these fluxes can break the GUT group down to that of the MSSM while lifting some of the charged exotics that descend from the 8-dimensional adjoint vector multiplet.

Most of the recent literature has focused on models in which the internal four-cycle wrapped by the GUT 7-branes is a del Pezzo surface $dP_n$, i.e. a surface obtained by blowing up $n$ points on  $\mathbb{P}^2$, $n=1, \cdots, 8$.  This was originally motivated by the desire to build models in which Planck scale physics could in principle be decoupled from that of the GUT \cite{Beasley:2008kw}.  Apart from this, however, del Pezzo's are promising for model building due to the fact that their bundle cohomologies exhibit a number of vanishing theorems \cite{Beasley:2008dc}.  In many cases, index formulae can be used to compute the precise spectrum of zero modes rather than net chiralities.  The vanishing of many cohomology groups also makes it easier to lift charged exotics after GUT breaking.

Nevertheless, it has been explicitly demonstrated that charged exotics cannot be completely removed by internal fluxes in $SO(10)$ models and this likely extends to higher rank groups as well.  While one can try to engineer models that utilize additional mechanisms to remove exotics, these difficulties seem to single out $SU(5)$ models, where exotic-free spectra can be obtained \cite{Beasley:2008kw}.

We will therefore focus on $SU(5)$ GUTs that utilize a nontrivial internal hypercharge flux, $F_{Y}$, to break the gauge group down to $SU(3)\times SU(2)\times U(1)_Y$.  Such a flux also projects out all light charged exotics that descend from the $SU(5)$ adjoint provided that its dual 2-cycle inside $S_{\rm GUT}$ is the difference of two exceptional classes of the del Pezzo\footnote{Throughout the paper, the generators of $H_2(dP_n, \mathbb{Z})$ will be denoted by $\tilde{h}$ (hyperplane class)  and $\tilde{e}_i$ (exceptional classes), with non-trivial intersections $\tilde{h}^2 =1$ and $\tilde{e}_i \cdot \tilde{e}_j =-\delta_{ij}$.}
\be
[F_Y]=\tilde{e}_i-\tilde{e}_j \,.
\ee
It is important to note, however, that this flux will generate a mass for the 4-dimensional hypercharge gauge boson unless the 2-cycles $\tilde{e}_i$ and $\tilde{e}_j$ are homologous to one another inside of the base $X_3$ of the elliptic fibration.  This is the only topological requirement we are aware of that both follows directly from phenomenological concerns in the local model and cannot be addressed in a local context.

In summary, the local input so far is as follows:
\begin{enumerate}
\item $SU(5)$ gauge group from 5 D7-branes wrapping a four-cycle $S_{\rm GUT}$
\item $S_{\rm GUT}$ is a del Pezzo surface, $dP_n$
\item At least two of the exceptional classes inside $S_{\rm GUT}=dP_n$ are homologous in the full geometry
\end{enumerate}


\subsection{Matter Content and Flavor Structure}

Since charged fields that descend from the $SU(5)$ adjoint are projected out by the hypercharge flux, the matter multiplets in F-theory GUTs arise from the intersection of the GUT branes with additional 7-branes.
In geometric language, this corresponds to complex codimension 1 surfaces inside $S_{\rm GUT}$ along which the singularity type of the elliptic fiber is enhanced in rank.  Charged fields in the $\mathbf{5}$ or $\mathbf{\overline{5}}$ arise when the enhancement is to $SU(6)$ while one obtains $\mathbf{10}$'s or $\mathbf{\overline{10}}$'s when the enhancement is to $SO(10)$.  The number of charged fields that one obtains along a given surface $\Sigma\subset S_{\rm GUT}$ is determined by bundle cohomology.  Important to note here is that one obtains a purely chiral spectrum whenever $\Sigma=\mathbb{P}^1$, with the number of zero modes determined by the net gauge flux along $\Sigma$.

Hypercharge flux $F_{Y}$ differentiates among the various components
of any GUT multiplet so a matter curve $\Sigma$ to which $F_{Y}$
restricts nontrivially will in general yield different numbers of
each.  It is therefore natural to engineer the usual MSSM matter
content, namely 3 $\mathbf{10}$'s and 3 $\mathbf{\overline{5}}$'s,
on matter curves to which $F_{Y}$ restricts trivially
\be
F_Y|_{\Sigma_{\mathbf{10}, M}} = F_Y|_{\Sigma_{\mathbf{\bar{5}}, M}} = 0 \,.
\ee
On the other
hand, we should engineer the $\mathbf{5}$ and
$\mathbf{\overline{5}}$ from which the Higgs fields descend on
matter curves where $F_{Y}$ restricts nontrivially.  In this case, a
suitable choice of gauge flux on the matter branes can project out
the Higgs triplets, leaving us with a single pair of doublets, $H_u$
and $H_d$.  When $H_u$ and $H_d$ originate on different
matter curves, this provides a realization of the missing partner
mechanism, hence evading current constraints from proton decay
experiments.

Yukawa couplings originate from isolated points where three matter curves come together inside $S_{\rm GUT}$.  Upon dimensional reduction, the couplings of light 4d fields are given at leading order by the product of their internal wave functions at the common point.  Because of this, the matrix of Yukawas from that originates from such a point has rank one at leading order so that it exhibits a single large eigenvalue \cite{Beasley:2008kw}.  Further, subleading corrections generate a natural hierarchical structure of the rough sort that is needed in the MSSM
\cite{Heckman:2008qa, Bouchard:2009bu,Randall:2009dw}.

If the same Yukawa coupling gets contributions from several such points, though, this nice structure can become distorted.  This is because these contributions typically cannot be simultaneously diagonalized at leading order.   For this reason, local models with nice hierarchies obtain the full matrix of $\mathbf{10}\times\mathbf{\overline{5}}\times\mathbf{\overline{5}}$ from a single, unique point where the corresponding matter curves intersect.  Similarly, one needs to obtain the full matrix of $\mathbf{10}\times\mathbf{10}\times\mathbf{5}$ from a single, unique point.  In the end, one generates the superpotential couplings
\be
W_{SU(5)} \supset  
\lambda_{\rm bottom}\mathbf{10}_M\times\mathbf{\overline{5}}_M\times\mathbf{\overline{5}}_{\bar{H}} +\lambda_{\rm top} \mathbf{10}_M\times\mathbf{10}_M\times\mathbf{5}_{H} \,.
\ee
 To achieve this, it is necessary to obtain all three generations of the $\mathbf{10}$ ($\mathbf{\overline{5}}$) from a single matter curve, denoted $\Sigma_{\mathbf{10}_M}$ ($\Sigma_{\mathbf{\overline{5}}_M}$).  
 Both Yukawa couplings will be generated on the same footing, from points of enhanced symmetry, namely $SO(12)$ for $\lambda_{\rm bottom}$ and $E_6$ for $\lambda_{\rm top}$. 


To obtain realistic matter content and hierarchical Yukawas, we thus require

\begin{enumerate}
\item $ 3 \times \mathbf{\overline{5}}$'s from a single matter curve,
$\Sigma_{\mathbf{\overline{5}}_M}$, with  $F_{Y}|_{\Sigma_{\mathbf{\overline{5}}_M}} =0$ 
\item $3 \times \mathbf{10}$'s from a single matter curve,
$\Sigma_{\mathbf{10}_M}$, with  $F_Y|_{\Sigma_{\mathbf{10}_M}} =0$
\item One doublet $H_u$ of  $\mathbf{5}_H$ from a single matter curve,
$\Sigma_{\mathbf{5}_H}$, with $F_Y|_{\Sigma_{\mathbf{5}_H}}\not =0 $
\item One doublet $H_d$ of  $\mathbf{\overline{5}}_{\bar{H}}$ from a single matter curve,
$\Sigma_{\mathbf{\overline{5}}_H}$, with $F_Y|_{\Sigma_{\mathbf{\overline{5}}_H}}\not=0$

\item A unique $E_6$ enhancement point where $\Sigma_{\mathbf{5}_M}\cap\Sigma_{\mathbf{10}_M}$ 
\item A unique $SO(12)$ enhancement point where $\Sigma_{\mathbf{10}_M}\cap \Sigma_{\mathbf{\overline{5}}_M} \cap \Sigma_{\mathbf{\overline{5}}_H}$
\end{enumerate}

Equipped with these phenomenological requirements we now turn to discuss the global consistency constraints for F-theory compactifications. 



\section{Building Compact Models}
\label{sec:Global}

We now turn to our basic strategy for building compact F-theory GUTs and a review of the various global constraints that they must satisfy.


\subsection{Elliptically Fibered Calabi-Yaus}
\label{subsec:EllipticCY}

To specify an F-theory compactification, we must construct an elliptically fibered Calabi-Yau (CY) four-fold $Y_4$ with base $X_3$ and describe the various $G$-fluxes on that four-fold.  In this subsection, we will focus on $Y_4$.

The main result of our work is the construction of a base three-fold $X_3$, which we will explain in section \ref{sec:ThreeFold}. This three-fold is by construction almost Fano, i.e.
\be
K_{X_3}^{-1}  \hbox{ is almost ample}
\,,
\ee
and one can follow standard procedures to construct an elliptically fibered CY four-fold with $X_3$ as base. Recall that an ample line bundle ${\cal{L}}$ is one for which some power ${\cal{L}}^n$ is very ample, which means that it has enough sections to construct a projective embedding of its base.

We now review the structure of local Calabi-Yau four-folds for $SU(5)$ GUTs  \cite{Andreas:2009uf}.  There, the four-fold is described as a local ALE fibration over a four-cycle, $S_{\rm GUT}$, over which the fiber degenerates to an $SU(5)$ singularity.  This is typically expressed in terms of a Weierstrass equation
\begin{equation}
x^2=y^3+fy+g\,,
\label{weierstrass}\end{equation}
where $f$ and $g$ are sections of suitable bundles over $S_{\rm GUT}$.  In order for this to describe a local Calabi-Yau geometry, $f$ must be a section of $\left(K_{S_{\rm GUT}}^{-1}\otimes N_{S_{\rm GUT}/X_3}\right)^4$ and $g$ a section of $\left(K_{S_{\rm GUT}}^{-1}\otimes N_{S_{\rm GUT}/X_3}\right)^6$.  To construct $f$ and $g$ with the right singularity structure, we identify a local coordinate $z$ normal to $S_{\rm GUT}$ and write $f$ and $g$ in a series expansion
\begin{equation}f = \frac{1}{2^4\cdot 3}\sum_m f_m z^m\,,\qquad g=\frac{1}{2^5\cdot 3^3}\sum_n g_nz^n\,,
\label{fgexpansion}\end{equation}
with $f_m$ and $g_n$ sections of the appropriate bundles.  In general, one can construct sections $f$ and $g$ with the desired structure by simply specifying suitable nonzero coefficient sections in this series.  For instance, to obtain an $SU(5)$ singularity along $S_{\rm GUT}$ with rank one enhancements to $SO(10)$ and $SU(6)$, one must choose \cite{Andreas:2009uf}
\begin{equation}\begin{split}f_0 &= -h^4 \\
f_1 &= 2h^2 H \\
f_2 &= 2hq-H^2\\
g_0 &= h^6 \\
g_1 &= -3h^4H \\
g_2 &= 3h^2(H^2-hq) \\
g_3 &= \frac{3}{2}h(2Hq-hf_3)-H^3 \\
g_4 &= \frac{3}{2}(f_3H+q^2) \,.
\label{fgcoeffs}\end{split}\end{equation}
For simplicity, we will assume $f_m=g_n=0$ for $m>3$ and $n>5$.  The structure of matter curves can now be obtained from the discriminant
\begin{equation}\Delta \sim z^5\left(h^4P + h^2\left[-2HP+hQ\right]z+\left[-3q^2H^3+{\cal{O}}(h)\right]+{\cal{O}}(z^2)\right)\end{equation}
with
\begin{equation}
\ba
P&=-3Hq^2-3hf_3q+2g_5h^2\cr
Q&= - q^3 - h \left({3\over 4} f_3^2  + 2 g_5 H \right) \,.
\ea
\label{PQdef}\end{equation}
Using the Kodaira classification, it is now easy to verify that rank one singularity enhancements arise as follows
\be
\ba
(h=0)\  & \Rightarrow  SO(10)\cr
(P=0) \ & \Rightarrow SU(6) \,.
\ea
\ee
Furthermore, we have the following rank two enhancements 
\be
\ba
(h=H=0) \ & \Rightarrow E_6 \cr
(h=q=0)\  & \Rightarrow  SO(12) \cr
(P=Q=0) \ & \Rightarrow  SU(7) \,.
\ea
\ee
These are precisely the loci where the Yukawa couplings are generated.

We can obtain an intuitive understanding for these singularity enhancements by performing a simple change of variables to the so-called Tate form \cite{Bershadsky:1996nh}.  Implicitly defining new sections $X$ and $Y$ as
\begin{equation}\begin{split}x&= X + \frac{1}{12}(h^2-Hz) \\
y&= Y-\frac{hX}{2}-\frac{qz^2}{24}\end{split}\end{equation}
our fibration takes the form
\begin{equation}Y^2 =X^3 + b_5 XY + b_4X^2z + b_3 Yz^2 + b_2 Xz^3 + b_0z^5\label{E8def}\,,
\end{equation}
where
\begin{equation} \label{Tate}
\ba
b_5 &= h \cr
b_4 &= -\frac{1}{4}H \cr
b_3 &= \frac{1}{12}q\cr
b_2 &= \frac{1}{48}f_3 \cr
b_0 &= \frac{1}{2^5\cdot 3^3}g_5\,.
\ea
\end{equation}

\begin{figure}
\begin{center}
\epsfig{file=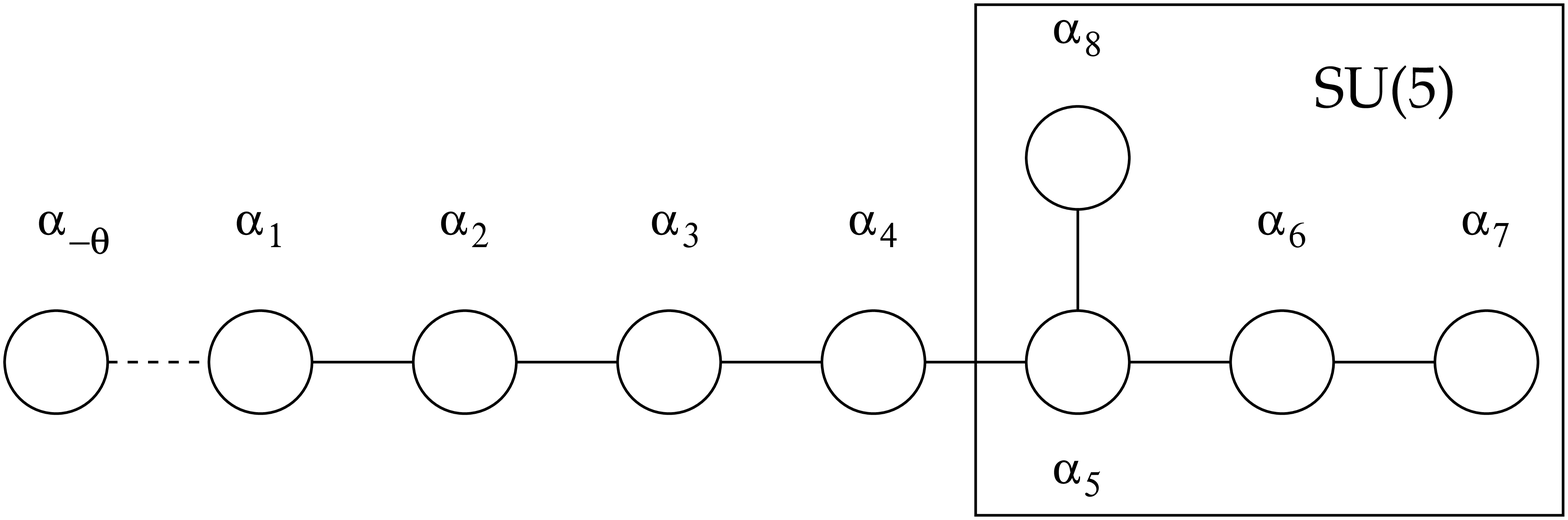,width=0.6\textwidth}
\caption{Extended $E_8$ Dynkin diagram}
\label{E8Dynkin}
\end{center}
\end{figure}

 This is nothing other than an $E_8$
singularity unfolded to $SU(5)$.  The geometry of this situation is
nicely described in \cite{Donagi:2008kj,Donagi:2009ra} and we briefly review it
here.  An $E_8$ singularity has 8 collapsed $\mathbb{P}^1$'s whose
intersection matrix is $-1$ times the Cartan matrix of $E_8$.  As
such, we can naturally identify them with nodes of the extended
$E_8$ Dynkin diagram, which we reproduce in figure \ref{E8Dynkin}.

To unfold this to $SU(5)$, the $\mathbb{P}^1$'s corresponding to $\alpha_1$, $\alpha_2$, $\alpha_3$, and $\alpha_4$ are resolved to nonzero volume.  Note also that the $\mathbb{P}^1$ corresponding to the additional node $\alpha_{-\theta}$ of the extended diagram,
\begin{equation}
\alpha_{-\theta} = -2\alpha_1-3\alpha_2-4\alpha_3-5\alpha_4-6\alpha_5-4\alpha_6-2\alpha_7-3\alpha_8\,,
\end{equation}
is also resolved here.  These $\mathbb{P}^1$'s are mixed under the subgroup $W_{A_4}$ of the Weyl group of $E_8$ which leaves $\alpha_5,\alpha_6,\alpha_7,\alpha_8$ fixed.

An enhancement to $SO(10)$ occurs when $\alpha_4$ or any of its images under $W_{A_4}$ shrinks to zero size.  There are five such $\mathbb{P}^1$'s, which we denote as
\begin{equation}\begin{split}\lambda_1 &= \alpha_4 \\
\lambda_2 &= \alpha_3+\alpha_4 \\
\lambda_3 &= \alpha_2+\alpha_3+\alpha_4 \\
\lambda_4 &= \alpha_1+\alpha_2+\alpha_3+\alpha_4 \\
\lambda_5 &= \alpha_{-\theta}+\alpha_1+\alpha_2+\alpha_3+\alpha_4\,.
\end{split}\end{equation}
In the following, we will abuse notation and use $\lambda_i$ to denote both a particular $\mathbb{P}^1$ as well as its volume.  As described in \cite{Donagi:2008kj,Donagi:2009ra}, the volumes $\lambda_i$ are encoded in the coefficients $b_n$ as
\begin{equation}b_n= b_0 s_n(\lambda_i)\,,
\end{equation}
where the $s_n(\lambda_i)$ are Schur polynomials.  Since an $SO(10)$ singularity occurs whenever any one of these $\mathbb{P}^1$'s shrinks to zero size, we see that the $SO(10)$ singular locus corresponds to solutions to $b_5\sim h=0$, as we expected.  On the other hand, an $SU(6)$ singularity occurs when $\lambda_i+\lambda_j\rightarrow 0$ for some $i\ne j$.  In terms of the $b_n$, this condition amounts to
$0=b_3^2b_4-b_2b_3b_5+b_0b_5^2=1728 P=0$.


Given a compact three-fold $X_3$ containing $S_{\rm GUT}$, it is trivial to uplift this general structure to an elliptic fibration over $X_3$.  We start again with the Weierstrass equation \eqref{weierstrass} but now with $f$ and $g$ suitable sections on $X_3$.  To obtain a four-fold that is Calabi-Yau, $f$ must be a section of $K_{X_3}^{-4}$ and $g$ a section of $K_{X_3}^{-6}$.  To construct $f$ and $g$, we identify a section $z$ of the bundle ${\cal{O}}(S_{\rm GUT})$ and specify $f$ and $g$ as an expansion in $z$ of the form \eqref{fgexpansion},\eqref{fgcoeffs}.  The bundles of which $z,h,H,q,f_3,g_5$ must be sections in order to define a local or global model are trivially related by adjunction $K_{S_{\rm GUT}}= K_{X_3}|_{S_{\rm GUT}}\otimes N_{S_{\rm GUT}/X_3}$.  For ease of reference, we list them in the table below
\begin{equation}\begin{array}{c|c|c} \text{Section} & \text{Bundle for Local Model} & \text{Bundle for Global Model} \\ \hline
z & N_{S_{\rm GUT}/X_3} & {\cal{O}}(S_{\rm GUT}) \\
h & K_{S_{\rm GUT}}^{-1}\otimes N_{S_{\rm GUT}/X_3} & K_{X_3}^{-1} \\
H & K_{S_{\rm GUT}}^{-2}\otimes N_{S_{\rm GUT}/X_3} & K_{X_3}^{-2}\otimes {\cal{O}}(-S_{\rm GUT}) \\
q & K_{S_{\rm GUT}}^{-3}\otimes N_{S_{\rm GUT}/X_3} & K_{X_3}^{-3}\otimes {\cal{O}}(-2S_{\rm GUT}) \\
f_3 & K_{S_{\rm GUT}}^{-4}\otimes N_{S_{\rm GUT}/X_3} & K_{X_3}^{-4}\otimes {\cal{O}}(-3S_{\rm GUT}) \\
g_5 & K_{S_{\rm GUT}}^{-6}\otimes N_{S_{\rm GUT}/X_3} & K_{X_3}^{-6}\otimes {\cal{O}}(-5S_{\rm GUT})\\
P  &   K_{S_{\rm GUT}}^{-8} \otimes N_{S_{\rm GUT}/X_3}^3 & K_{X_3}^{-8} \otimes {\cal{O}}(-5S_{\rm GUT}) \\
Q & K_{S_{\rm GUT}}^{-9} \otimes N_{S_{\rm GUT}/X_3}^3 & K_{X_3}^{-9} \otimes {\cal{O}}(-6S_{\rm GUT})
\end{array}\label{localglobalbunds}\end{equation}
In passing from a global model of this sort back to a local one, we can use $z$ to define a local coordinate in a neighborhood of $S_{\rm GUT}$ and further expand the sections $h,H,q,f_3,g_5$ in $z$.  From \eqref{E8def}, it is clear that terms beyond the leading order in such an expansion do not affect the singularity structure along $S_{GUT}$.

It goes without saying that we need all of the bundles in \eqref{localglobalbunds} to admit holomorphic sections.  
The three-fold $X_3$ that we construct later will have the property that each of these bundles corresponds to an effective divisor so the existence of holomorphic sections will be guaranteed{\footnote{In fact, we will give explicit formulae for all of the various holomorphic sections.}}.


\subsection{Hypercharge Flux condition}

As discussed in section \ref{subsec:GUTbreaking}, the GUT group is broken  by switching on hypercharge flux along
$S_{GUT}$ which lifts the triplets of the Higgs ${\bf 5}$ and
$\overline{\bf 5}$ multiplets.  For $S_{GUT}=dP_n$, this can be accomplished with a flux of the form \cite{Beasley:2008kw,Donagi:2008kj}
\be 
[F_Y] = \tilde{e}_i - \tilde{e}_j \,. 
\ee 
where $\tilde{e}_i$ and $\tilde{e}_j$ are two exceptional curves in $dP_n$.  In order for the
$U(1)_Y$ gauge boson to remain massless, these two curve classes must be equivalent as elements of $H_2(X_3,\mathbb{Z})$ \cite{Buican:2006sn}.  In particular, there has to exist a three-chain
$\Omega_3$ in $X_3$, such that 
\be\label{HyperConst}
\partial \Omega_3 = {\tilde{e}_i} \cup  \left(-\tilde{e}_j \right) \,.
\ee
This is an important nontrivial constraint that the  three-fold has to satisfy.


\subsection{Quantization of $G$-flux and Tadpole Constraints}

As familiar from type II intersecting brane models, a number of global consistency conditions must be satisfied.  These include tadpole cancellation constraints as well as cancellation of the Freed-Witten anomaly where applicable.  We now briefly discuss the analogs of these constraints in F-theory models.

\subsubsection{Quantization of $G$-Fluxes}

First, let us address the F-theory analog of the Freed-Witten anomaly.  In intersecting brane constructions, this anomaly arises on string worldsheets with boundary unless worldvolume fluxes on the corresponding branes are suitably quantized \cite{Freed:1999vc}.  A similar discussion of anomalies in the $M2$ membrane theory with boundary \cite{Witten:1996md} implies that the $G$-flux on $M$-theory backgrounds, and hence also their F-theory duals, must obey the quantization condition
\begin{equation}
\left[G_4\right] - \frac{c_2(Y_4)}{2}\in H^4(Y_4,\mathbb{Z})\,.
\label{Gfluxquant}\end{equation}
For generic Calabi-Yau four-folds, $Y_4$, there is no reason for $c_2(Y_4)$ to be an even class so it may be necessary to turn on nontrivial half-integral $G$-fluxes to satisfy this constraint.  Because some $G$-fluxes break 4-dimensional Lorentz invariance and others capture nontrival field strengths on various brane worldvolumes, it is important to understand which, if any, $G$-fluxes must be half-integrally quantized to satisfy \eqref{Gfluxquant}.  

For F-theory compactifications on an elliptically fibered $Y_4$ with smooth Weierstrass form, one can use the approach of \cite{Sethi:1996es} to study $c_2(Y_4)$.  For this, we realize $Y_4$ as a suitably ``homogenized" Weierstrass equation of the form
\begin{equation}
s=ZY^2-X^3+aXZ^2-bZ^3=0\label{homoweier}\,,
\end{equation}
so that $X,Y,Z$ can be thought of as homogenous coordinates on a $\mathbb{P}^2$ bundle $W\rightarrow X_3$.  These coordinates are sections of ${\cal{O}}(1)\otimes K_{X_3}^{-2}$, ${\cal{O}}(1)\otimes K_{X_3}^{-3}$, and ${\cal{O}}(1)$, respectively, where ${\cal{O}}(1)$ is a line bundle on $W$ that restricts to a degree 1 line bundle on each $\mathbb{P}^2$ fiber.  The cohomology ring of $W$ is generated by the cohomology ring of $X_3$ along with the class $\alpha=c_1({\cal{O}}(1))$ that descends from the $\mathbb{P}^2$ subject to the relation
\begin{equation}
\alpha(\alpha+2c_1(X_3))(\alpha+3c_1(X_3))=0\,,
\label{Wequiv}\end{equation}
which follows from emptiness of the intersection $X=Y=Z=0$.  The total Chern class of $Y_4$ can now be obtained by adjunction, leading to the following expression for $c_2(Y_4)$
\begin{equation}
c_2(Y_4)=11c_1(B_2)^2+c_2(B_2)+13c_1(B_2)\alpha+3\alpha^2\,.
\label{c2Y4}\end{equation}
Because $s$ in \eqref{homoweier} is a section of $c_1({\cal{O}}(1))\otimes K_{X_3}^{-6}$, any integration inside $Y_4$ can be extended to an integration inside $W$ provided we multiply the integrand by $3(\alpha+2c_1(X_3))$.  Because of \eqref{Wequiv}, this means that we have an additional equivalence relation on $Y_4$ that does not extend to all of $W$
\begin{equation}\alpha(\alpha+3c_1(X_3))=0\,,
\label{WWequiv}\end{equation}
which allows us to write $c_2(Y_4)$ as
\begin{equation}c_2(Y_4)=11c_1(X_3)^2 + c_2(X_3)+4\alpha c_1(X_3)\,.
\label{c2Y4red}\end{equation} 
The integral of $c_2(Y_4)$ over any four-cycle that includes part of the elliptic fiber will get contributions only from the $\alpha$-dependent term in \eqref{c2Y4red} so the result is always even.  Note that this includes all four-cycles on which we can add $G$-fluxes without breaking 4-dimensional Lorentz invariance.  

It remains to study integrals of $c_2(Y_4)$ over four-cycles contained entirely within the base, $X_3$.  For this, we must be careful to accurately account for the contribution of $\alpha$, which can have a nontrivial restriction to $X_3$.  This can be done by noting that restricting to $X_3$ amounts to setting $Z=0$ or, equivalently, to multiplying the integrand by $\alpha$.  Recalling \eqref{Wequiv} and \eqref{WWequiv}, this means that $\alpha$ and $c_1(B_2)$ satisfy an additional equivalence relation inside $X_3$
\begin{equation}\alpha+3c_1(X_3)=0\,,
\end{equation}
so that when $c_2(Y_4)$ is restricted to $X_3$, it takes the form
\begin{equation}c_2(Y_4)|_{X_3} = c_2(X_3) - c_1(X_3)^2\,.
\end{equation}
This is precisely what we expect for the base of an elliptically fibered Calabi-Yau from the standard adjunction formula.  For the specific $X_3$ that we shall construct later, it will be easy to directly compute $c_2(X_3)$ and $c_1(X_3)$ in order to verify that this is an even class.  It is possible that $c_2(X_3)-c_1(X_3)^2$ is an even class for generic almost Fano three-folds $X_3$ but we are currently unaware of a theorem to this effect.

It is important to note that, strictly speaking, these arguments apply only to elliptic fibrations with smooth Weierstrass form.  By contrast, for model building we would like to consider elliptically fibered four-folds whose Weierstrass form is not smooth due in part to the $SU(5)$ degeneration locus on which the GUT gauge group is localized.  We are therefore making an implicit assumption, namely that the relevant object in the quantization condition \eqref{Gfluxquant} for our $Y_4$ is the second Chern class of the smooth four-fold obtained upon resolution of all singularites. 

\subsubsection{Tadpole Conditions}

Any consistent compactification must also satisfy a number of tadpole cancellation conditions.  We briefly discuss each of these in the context of F-theory compactifications

As is well known, 7-brane tadpole cancellation in F-theory follows immediately from the Calabi-Yau condition{\footnote{See, for instance, \cite{Andreas:2009uf}.}}.  
Cancellation of the D5-brane tadpole, on the other hand, follows from primitivity of the $G$-flux
\be \label{prim} G_4\wedge J_{X_3}=0, \ee
which ensures supersymmetry of the background. Here, $J_{X_3}$ is
a K\"ahler form on $X_3$.

This leaves us only with the D3-brane tadpole.  In most examples, it is expected that the induced D3-brane charge of an F-theory compactification can be computed with the Euler character $\chi(Y_4)$.  This has in fact been checked in several examples with heterotic duals{\footnote{This includes examples with singular $Y_4$ which must first be resolved in order to compute $\chi$ \cite{Andreas:2009uf}.  It should be noted, however, that a mismatch was found for a particularly ``badly behaved" example in \cite{Andreas:2009uf}.  This example amounted to completely turning off the deformation $h$ in \eqref{fgcoeffs} so we expect that for more generic models this will not be a problem.}}.  In what follows, we will assume that it holds also for our $Y_4$ so that, including the $G$-flux contribution, cancellation of the D3-brane tadpole amounts to imposing
\begin{equation}
\frac{\chi(Y_4)}{24} = \frac{1}{2}\int_{Y_4}\,G_4\wedge G_4 + N_{D3} \,.
\label{D3tad}
\end{equation}
Two important concerns regarding the D3 tadpole are its integrality and its sign.  Firstly, we expect that the number $N_{D3}$ of D3-branes that must be introduced to cancel the tadpole is an integer.  Secondly, we would like $N_{D3}$ to be positive so that it does not become necessary to introduce any $\overline{\text{D3}}$-branes.  While such $\overline{\text{D3}}$-branes might provide a candidate mechanism for breaking supersymmetry, we will not consider this possibility here.

To address integrality of the D3-brane tadpole in earnest, we must study not only $\chi(Y_4)$ but take into account all $G$-fluxes that are present.  In this paper, we will not embark on a discussion of $G$-fluxes in full generality.   The cancellation of any half-integral contributions to \eqref{D3tad} from $G$-fluxes or $\chi(Y_4)$, then, will remain an important input constraint for the study of moduli stabilization in these backgrounds.

Nevertheless, it remains important that $\chi(Y_4)/24$ be at least half-integral.  Using the techniques reviewed above, the authors of \cite{Sethi:1996es} demonstrated that the Euler character of a smooth elliptically fibered four-fold $Y_4$ can be expressed as an integral over Chern classes of the base, $X_3$, through a formula
\begin{equation}\label{D3tadsimp}
\chi (Y_4) = 12\int_{X_3}\,c_1(X_3)\left(c_2(X_3)+30c_1(X_3)^2\right) \,,
\end{equation}
which makes this fact manifest.  Of course, the four-folds that we need for model building do not admit a smooth Weierstrass form but we again assume that the relevant object for \eqref{D3tad} is the Euler character of the smooth four-fold obtained from a suitable resolution of $Y_4$.

\section{Construction of the Three-fold Base}
\label{sec:ThreeFold}

We now turn to the main object of this paper, namely constructing a family of three-folds which can serve as the base for elliptically fibered Calabi-Yau four-folds for $SU(5)$ GUTs.
The essential criterion that we impose is that such a three-fold must contain a del Pezzo surface with a pair of exceptional curve classes that satisfy the hypercharge constraint \eqref{HyperConst}.

In section \ref{subsec:BasicIdea} we outline the strategy of our construction. We present a local description of the geometry in detail in section \ref{subsec:LocalConstruction}.  We then embed this setup in $\mathbb{P}^3$ to obtain an almost Fano three-fold, $X_3=\tilde{X}$, in section \ref{subsec:CompactBF}.  This three-fold contains a $dP_2$ surface satisfying \eqref{HyperConst} which can be blown up to $dP_n$ if desired.  Details of the geometry of $\tilde{X}$ are summarized in \ref{subsec:SummaryBase}.


\subsection{Basic Idea of the Construction}
\label{subsec:BasicIdea}

The starting point of our construction is the specification of a nodal curve inside $\mathbb{P}^3$.  Via a series of blow-ups, we obtain from this a smooth three-fold $X$.  Our final three-fold, $\tilde{X}$, is then obtained from this via a flop transition.  We will determine effective divisors for $X$ and $\tilde{X}$, compute their intersection numbers, and give explicit holomorphic representatives which provide us with the necessary holomorphic sections for constructing elliptic fibrations.

This type of three-fold has appeared in the context of Mori's Minimal Model Program for three-folds, in the
seminal paper \cite{Mori}, (3.44.2),  (see also \cite{Matsuki} for a concise summary of the mathematical setup, and \cite{Buican:2006sn} for a discussion in the context of hypercharge flux). The reverse operations to the blowups that we will describe are the so-called extremal contractions.
We shall, however, give a construction in fairly basic terms, by providing explicit details of the constructions, and thus no prior understanding of \cite{Mori} will be required. \\


\begin{center}
\begin{figure}
\hspace{1cm}
\epsfig{file=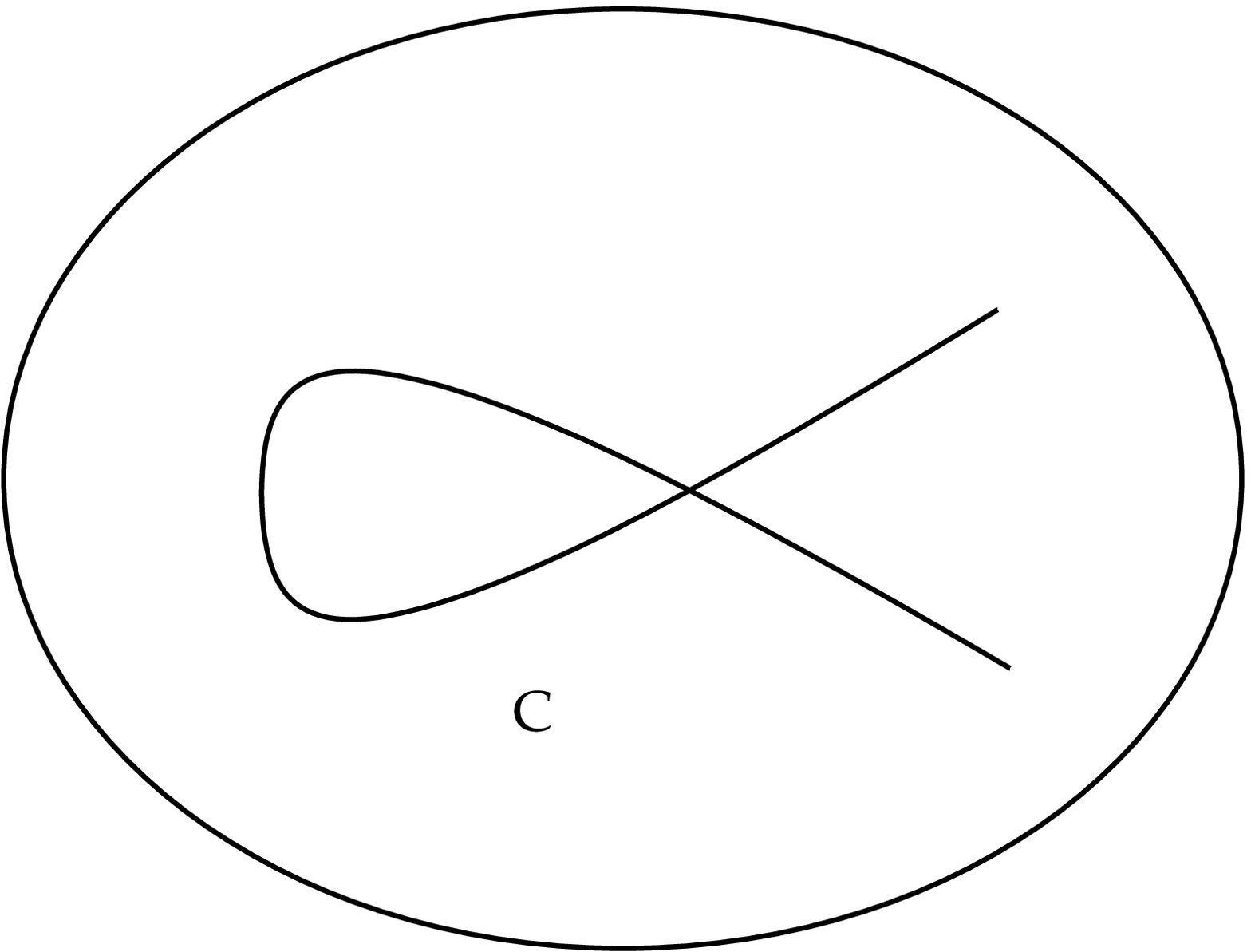,width=4.2cm}
\hspace{1.5cm}
\epsfig{file=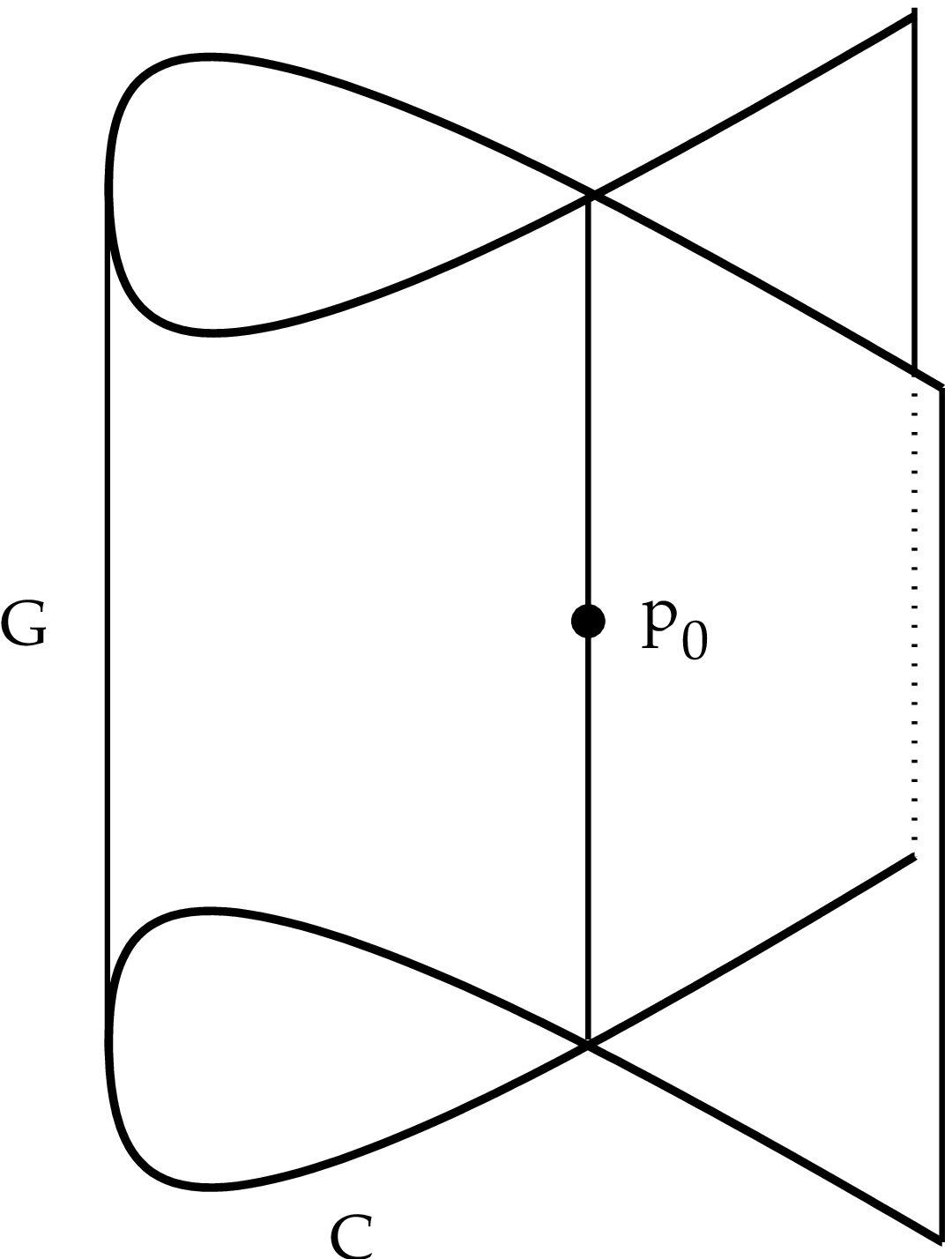,width=3.3cm }
\hspace{1.5cm}
\epsfig{file=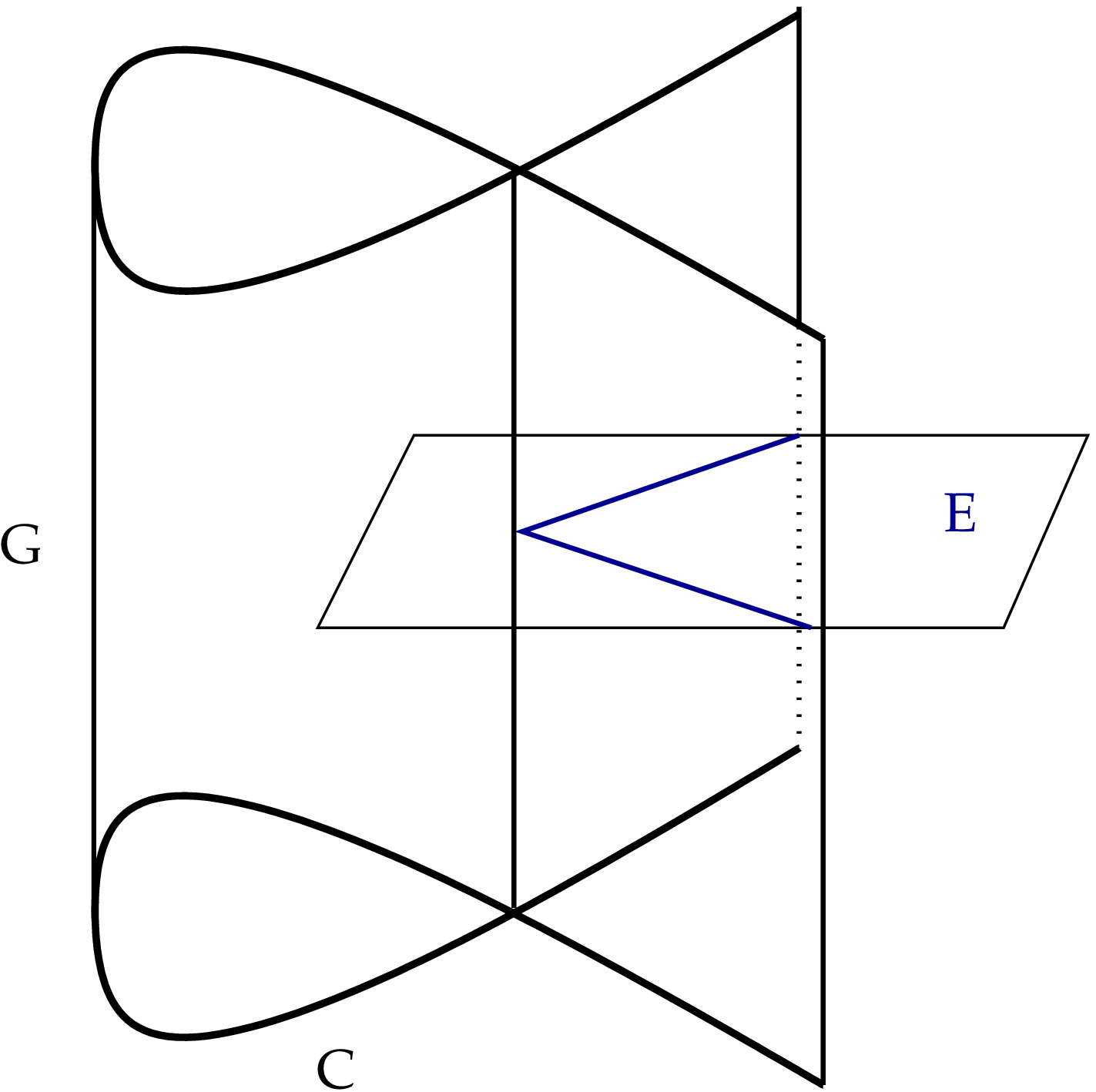,width=4.3cm }
\caption{Construction of the base three-fold:
The starting point is the nodal curve $C$ in $Z$. Blow-up along $C$ results in a new divisor with singular point $p_0$. Blowing up the point $p_0$ yields an exceptional divisor $E\equiv \mathbb{P}^1 \times \mathbb{P}^1$, where the two $\mathbb{P}^1$s are homologous in the resulting threefold. }
\label{fig:Mori}
\end{figure}
\end{center}


The setup is summarized  in figure \ref{fig:Mori}:

\begin{itemize}
\item  Choose a nodal curve $C$ in a three-fold $Z$ (which will be $\mathbb{P}^3$ for our purposes), and blow up along this curve. This results in a new three-fold $Y$, depicted in the middle of figure \ref{fig:Mori}. There is a singular point $p_0$, which lies over the nodal point of $C$.  This singular point is in fact a conifold singularity.

\item Blowing up the conifold singularity  at $p_0$ results in a smooth three-fold $X$.  This introduces an exceptional divisor $E$, which is isomorphic to $\mathbb{P}^1 \times \mathbb{P}^1$.  The two $\mathbb{P}^1$'s inside $E$, $\ell_1$ and $\ell_2$, are distinct in $H_2(E,\mathbb{Z})$ but are homologous inside $X$.  This is precisely the situation that we require in order to satisfy the hypercharge flux condition \eqref{HyperConst}.

\item The vertical curve $G$ is a $(-1, -1)$ curve (with normal bundle   
$\mathcal{O}(-1)\oplus \mathcal{O}(-1)$ in $X$), and can be flopped.  From the perspective of $E$, the effect of the flop is to blow up a point.  As such, $E$ transforms to a del Pezzo $dP_2$ whose exceptional curves are homologous in the three-fold.  This geometry is the three-fold  $\tilde{X}$ that we will use for the base of the elliptically fibered Calabi-Yau.

\end{itemize}

After the flop, we can further blow up points in $\tilde{X}$ in such a way that the $dP_2$ is increased to $dP_n$ with $n>2$.  We expect that this will be useful in the future for constructing models, which in addition to the GUT sector include e.g. a supersymmetry-breaking sector \cite{Marsano:2008jq}.


\subsection{Local Construction of the Three-fold}
\label{subsec:LocalConstruction}

As a first step we describe locally the various blow-ups leading to the construction of $\tilde{X}$. In the next section, we embed this into $\mathbb{P}^3$ and provide divisors and intersection numbers.





\begin{figure}
\begin{center}
\epsfig{file=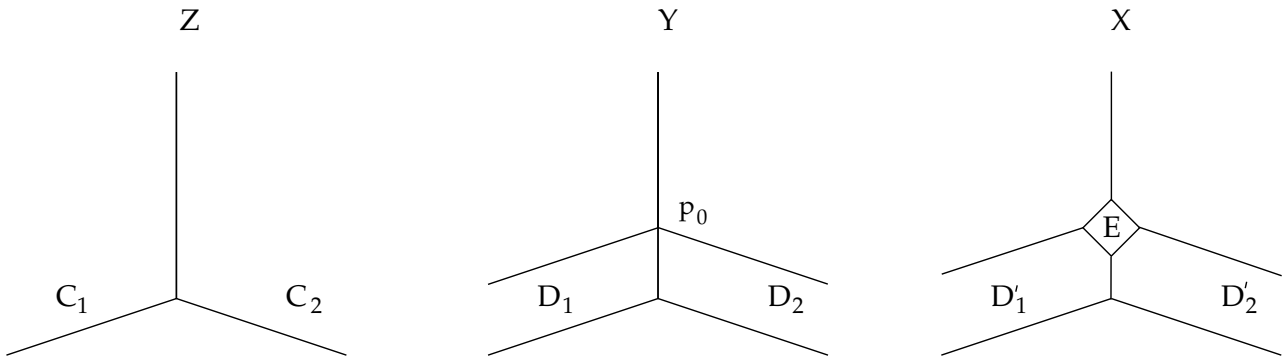,width=1\textwidth}
\caption{Local construction of the three-fold: blow-ups.}
\label{fig:LocalThreefold}
\end{center}
\end{figure}

For a local description the three-fold $Z$ can be approximated by $Z= \mathbb{C}^3$ and the nodal curve is described by
\be\label{LocalSing}
C\, :\quad x y = z=0  \,.
\ee
We see that $C$ has two components, $C = C_1 \cup C_2$, corresponding to $x=z=0$ and $y=z=0$.  Each component is isomorphic to $\mathbb{C}$ and the nodal point $p=(0,0,0)$ is the intersection $C_1 \cap C_2$, see figure \ref{fig:LocalThreefold}.

The first step requires blow-up along the curve $C$, so that the three-fold $Y$ locally takes the form
\be
Y = \left\{ ((x,y,z),[V_0, V_1]) \in \mathbb{C}^3 \times  \mathbb{P}^1_V: \quad xy V_0 = z V_1 \right\} \,.
\ee
The kernel of the blow-down map
\be
\psi\, : \quad  Y \rightarrow Z \,,
\ee
by definition contains a new divisor
\be
Q = \psi^{-1} (C) = D_1 \cup D_2 \,,\qquad  D_i = C_i \times \mathbb{P}^1_V \,,
\ee
which can be written in terms of the local coordinates as
\be
Q = \left\{ ((x,y,z),[V_0,V_1])\in  \mathbb{C}^3 \times  \mathbb{P}^1_V: \
        xy=z=0\right\} = \mathbb{C} \times  \mathbb{P}^1_V\,.
\ee
$Y$ can be covered by two patches, containing the north and south pole, respectively, of the $\mathbb{P}^1_V$
\be
Y_0 = \{V_0\neq 0\}  \qquad \hbox{and}\qquad Y_1 =\{V_1\neq 0\}\,.
\ee
In local coordinates on $\mathbb{C}^3 \times \{V_1 \not= 0\}$, which we denote by $x$, $y$, $z$, $t= V_0/V_1$
\be
Y_1=\left\{ (x,y,z,t) \in \mathbb{C}^4\, :\  xy t  = z\right\} \,.
\ee
Likewise on  $\mathbb{C}^3 \times \{V_0 \not= 0\}$ the local coordinates are
$x$, $y$, $z$, $u= V_1/V_0$, and
\be\label{LocCon}
Y_0=\left\{ (x,y,z,u) \in \mathbb{C}^4\, :\   xy = zu \right\} \,,
\ee
which exhibits a conifold singularity at the point
\be
p_0 = \{x=y=z=u=0 \} \,.
\ee


The three-fold $X$ is obtained by blowing up the conifold singularity,
i.e. by gluing in an exceptional divisor.  This can be done in the local patch $Y_0$ by first considering the $\mathbb{C}^4$ parametrized by $(x,y,z,u)$ and blowing up the origin by the standard method of gluing in a $\mathbb{P}^3$.  We then restrict to the conifold $xy=zu$ and take a smooth continuation as we approach the origin.  To blow up the origin of $\mathbb{C}^4$, we introduce a $\mathbb{P}^3_W$ with homogeneous coordinates $[W_1,W_2,W_3,W_4]$ and restrict to the submanifold of $\mathbb{C}^4\times\mathbb{P}^3_W$ obtained by imposing the condition that $(x,y,z,u)$ be contained inside the line $[W_1,W_2,W_3,W_4]$.  In equations, this statement is equivalent to
\be\label{ZinW}
\ba
z_iW_j &= z_j W_i \\
z_i W_4 &= u W_i\,,
\ea
\ee
for $i,j=1,2,3$.  Restricting to $xy=zu$ away from the origin further imposes $W_1W_2=W_3W_4$, which we retain as a condition on the full blown-up three-fold.  We therefore impose both equations
\be\label{WWeq}
z_1 z_2 = z_3 u \,,\qquad
W_1 W_2 = W_3 W_4 \,.
\ee
so that the three-fold takes the following form in this patch
\be
\ba
X_0 = &\left\{ ((x,y,z,u), [W_1,W_2,W_3,W_4])\in \mathbb{C}^4 \times  \mathbb{P}^3_W\,:\   \right.\cr
& \qquad \qquad\qquad \qquad\left. (x,y,z,u) \in [W_1,W_2,W_3,W_4]\,, \ xy=zu\,,\   W_1W_2=W_3W_4 \right\} \,.
\ea
\ee
The exceptional divisor of the blowdown map
$
\phi_0: X_0 \rightarrow Y_0
$
is
\be
E=\phi_0^{-1}(q)= \{(0,0,0,0)\} \times  \{[W_1,W_2,W_3,W_4] \in \mathbb{P}^3\,:\ W_1W_2=W_3W_4\}
 \cong \mathbb{P}^1 \times  \mathbb{P}^1\,.
\ee
The divisor $Q$ maps under the blow-up to a divisor in $X_0$
\be
\ba
Q_0 =&\big\{((x,y,z,u),\, [W_1,W_2,W_3,W_4]) \in \mathbb{C}^4 \times \mathbb{P}^3\,:\cr
&\qquad \qquad\quad  (x,y,z,u) \in [W_1,W_2,W_3,W_4]\,, \  xy=z=0\,,\   W_1W_2=W_3=0\big\} \,.
\ea
\ee
%
%


\begin{figure}
\begin{center}
\epsfig{file=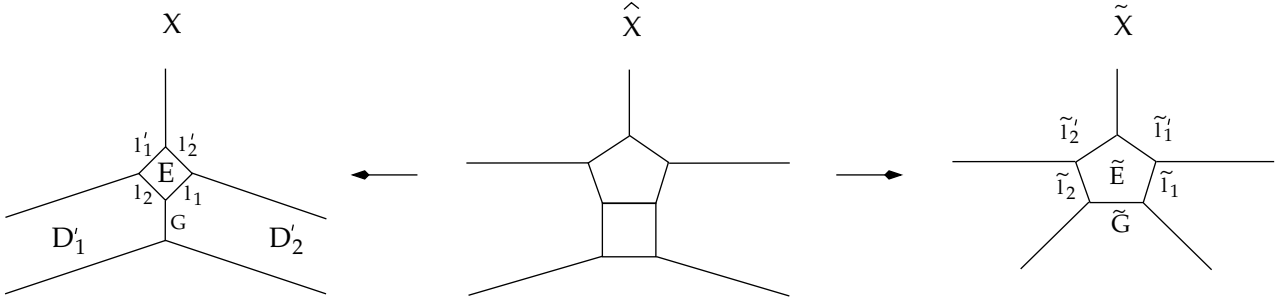,width=1\textwidth}
\caption{Local construction of the three-fold: flop transition between $X$ and $\tilde{X}$.}
\label{fig:LocalFlop}
\end{center}
\end{figure}




The exceptional divisor $E$ will house our GUT gauge group so we would like to know if it is possible to equip it with some more structure.  A simple way to achieve this is to note that the curve $G = D_1' \cap D_2'$ has normal bundles
\be
N_{G/D_i'} = \mathcal{O}(-1) \,,
\ee
and therefore can be flopped: i.e. we can blow this curve down, and blow-up another curve $\tilde{G}$, as depicted in figure \ref{fig:LocalFlop}. All other curves $\ell_i, \ell_i'$ are $(0, -1)$ curves.
After the flop we obtain the threefold $\tilde{X}$.  The flop effectively blows up a point in the divisor $E=\mathbb{P}^1\times\mathbb{P}^1$ so that it transforms into 
\be
\tilde{E} \cong  dP_2 \,,
\ee
When we realize this in a compact three-fold, $\tilde{X}$, the two exceptional classes of this $dP_2$ will be equivalent in $H_2(\tilde{X},\mathbb{Z})$ so that they may be used to construct a hypercharge flux for breaking the GUT gauge group.


\subsection{Embedding into $\mathbb{P}^3$: Before the Flop}
\label{subsec:CompactBF}

Now that we have explained the local geometry, we will embed the construction of the last subsection into a compact three-fold $Z$, which we take to be $\mathbb{P}^3$ for simplicity. This results in a three-fold $X$, which has a local patch given by the construction in section \ref{subsec:LocalConstruction}. We discuss all the relevant divisor classes and intersection numbers. In the next section, the final three-fold will be discussed, which is obtained from $X$ by a flop-transition.



\begin{figure}
\begin{center}
\epsfig{file=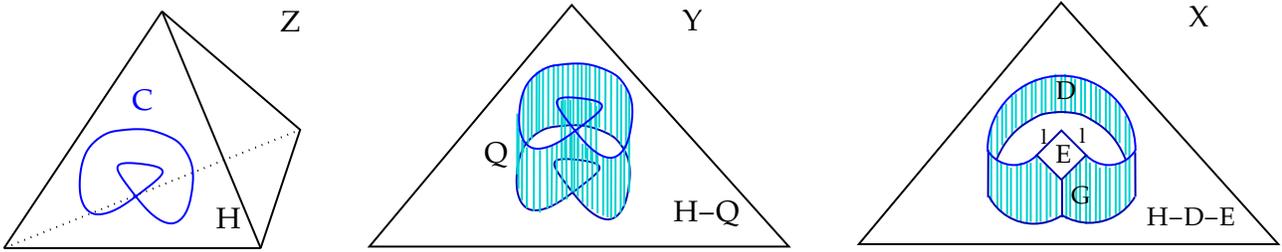,width=1\textwidth}
\caption{Global Construction of Threefold: blowups.}
\label{fig:Global12}
\end{center}
\end{figure}


\subsubsection{Construction of  the Three-fold}

Let $Z=\mathbb{P}^3$ with homogenous coordinates $[Z_0,Z_1,Z_2,Z_3]$.  The canonical class is given in terms of the hyperplane class $H$ as
\be
K_Z = - 4 H \,.
\ee
Inside $\mathbb{P}^3$, we consider the nodal curve ${\cal{C}}$ defined by the equations
\be
\ba
Z_4Z_1Z_2+(Z_1+Z_2)^3&=0 \cr
            Z_3 &=0\,.
\ea
\ee
Alternatively, this can be written in affine coordinates $z_i$ as
\begin{equation}
{\cal{C}}=\left\{[z_1,z_2,0,1]\,|\, z_1z_2+(z_1+z_2)^3=0\right\}\cup\left\{[1,-1,0,0]\right\}\,.
\label{Ceqns}\end{equation}
In what follows, we will typically consider the affine patch $[z_1,z_2,z_3,1]$ of $\mathbb{P}^3$ since this contains all of ${\cal{C}}$ except for a single ``point at infinity".  As clear from \eqref{Ceqns}, ${\cal{C}}$ exhibits a singular point at $[0,0,0,1]$ which is precisely of the form $z_1z_2=z_3=0$ as in the local description (\ref{LocalSing}).

The first step in constructing our three-fold is again to blow up along ${\cal{C}}$ to obtain the three-fold $Y$ with the blow-down map
\be
\psi: Y \rightarrow Z \,.
\ee
In coordinates this can be described by considering $\mathbb{C}^3\times\mathbb{P}^1$ in the $Z_4=1$ patch with homogeneous coordinates $[V_0,V_1]$ on the new $\mathbb{P}^1$, which we shall hereafter denote by $\mathbb{P}^1_V$.  The blow-up is then defined in this patch by the equation
\begin{equation}
Y\,:\qquad V_0\left(z_1z_2+(z_1+z_2)^3\right)=V_1z_3 \,.
\label{blowup1}
\end{equation}
From \eqref{blowup1}, we see that the resulting three-fold exhibits a singular point at $\{(z_1,z_2,z_3),[V_0,V_1]\}=\{(0,0,0),[1,0]\}$.  As in the local description (\ref{LocCon}), we pass to an
affine patch covering the north pole $v_0\not=0$ of $\mathbb{P}^1_V$. Then defining again $u = v_1 /v_0$ the equation \eqref{blowup1} in fact becomes
\begin{equation}
[z_1z_2+(z_1+z_2)^3]=uz_3\,,
\end{equation}
so that near the singular point it behaves like
\begin{equation}
z_1z_2=u  z_3\,.
\end{equation}
We recognize this as a conifold singularity.

The divisor classes in $Y$ are the exceptional divisor $Q$, which is a $\mathbb{P}^1$-bundle over ${\cal C}$ (more details on the precise geometry of $Q$ can be found in appendix \ref{app:Geo}), and $\psi^* (H)  = Q +(H-Q)$. The canonical class is
\be
K_Y = \psi^* (K_Z) + Q = - 4 H + Q \,.
\ee

The final step is to blow-up the conifold singularity in $Y$ by
\be
\phi: X\rightarrow Y \,.
\ee
To do this, we move to a local patch covering the north pole of $\mathbb{P}^1_V$ with coordinates $(z_1,z_2,z_3,u=v_1/v_0)$ and proceed exactly as in the local setup described in section \ref{subsec:LocalConstruction}.  In particular, we blow up the origin of this $\mathbb{C}^4$ by gluing in a $\mathbb{P}^3_W$ with homogeneous coordinates $[W_1,W_2,W_3,W_4]$ and restrict to $z_1z_2=z_3u$ and its smooth continuation, $W_1W_2=W_3W_4$, at the origin.  In the end, the three-fold takes the following form in this local patch
\be\label{XDef}
\ba
X_1 = & \left\{(z_1,z_2,z_3,v_1;W_1,W_2,W_3,W_4)\in \mathbb{C}^4\times \mathbb{P}^3_W\,:\quad \right.\cr
& \left.  \qquad \qquad
(z_1,z_2,z_3, u) \in [W_1,W_2,W_3,W_4] \,, \quad
 z_1 z_2 = z_3  u \,,\quad W_1 W_2 = W_3 W_4
  \right\}\,.
  \ea
\ee
As explained in section \ref{subsec:LocalConstruction}, the condition $(z_1,z_2,z_3, u) \in [W_1,W_2,W_3,W_4]$ can be made more explicitly written as in \eqref{ZinW}.

We can identify the two $\mathbb{P}^1$'s with the submanifolds
\be\label{P1s}
\mathbb{P}^1_{(1)}:\quad W_2=W_4=0\,,\qquad
\mathbb{P}^1_{(2)}:\quad W_2=W_3=0\,.
\ee
Note that in this local patch it is not possible to see that these $\mathbb{P}^1$s are in the same class in $X$.
It is however clear from the global topology of $X$ since their intersections with all divisors are equivalent.
The canonical class of $X$ is
\be
K_X = - 4 H + (D + E )+  E \,,
\ee
where the exceptional divisor is
\be
\phi^* Q = D + E  \,.
\ee


\subsubsection{Curves and Intersection numbers}

A detailed analysis and derivation of the intersection tables will be given in Appendix \ref{app:Ints} and in this section we only summarize the results.  As a basis of $H_2(X,\mathbb{Z})$, we take the curve $\ell_0$, which descends from the unique generator of $H_2(\mathbb{P}^3,\mathbb{Z})$, as well as the curves $\ell$ and $G$ depicted in figure \ref{fig:Global12}.
The intersection numbers with various divisors are given by the following table
\begin{center}
\begin{tabular}{|c||r|r|r|r|}
\hline
        &$H$    & $D$   &   $E$  \cr\hline\hline
$\ell_0$    & $+1$& $0$& $0$ \cr\hline
$\ell$      & $0$&$ +1$ & $-1$ \cr\hline
$G$     & $0$& $-2$ & $1$  \cr\hline
\end{tabular}
\end{center}
The intersections of divisors with one another is furthermore
\begin{equation}\label{DoubleIntBF}
\begin{array}{|c||c|c|c|}
\hline
& H & E & D \\ \hline\hline
H & \ell_0 & 0 & 3(\ell+G) \cr\hline
E & 0 & -2\ell & 2\ell \cr\hline
D & 3(\ell+G) & 2\ell & -3\ell_0 + 12(\ell+G) - 2\ell \cr\hline
H-D-E & \ell_0-3(\ell+G) & 0 & 3\left(\ell_0-3(\ell+G)\right)\cr\hline
\end{array}
\end{equation}
from which the following non-vanishing triple-intersections follow
\be
\ba
H^3 &= 1 \cr
D^3 &= -14 \cr
E^3 &= 2 \cr
D^2H &= -3 \cr
D^2E &= 2 \cr
E^2 D &= -2 \,.
\ea
\ee
We can confirm these intersection numbers from direct computations using the explicit description of divisors in the next subsection.


\subsubsection{Effective Divisors and Holomorphic Sections}
\label{subsec:DivX}

A basic tool for model-building will be the set of divisors, as these will provide us with holomorphic sections that we can use to construct elliptic fibrations.  We will present explicitly the holomorphic sections that define various effective divisors which are linear combinations of $H$, $D$, and $E$. The coordinates $Z_i, V_j$ and $W_k$ defined in (\ref{blowup1},\ref{XDef}, \ref{ZinW}, \ref{WWeq}) are the building blocks to write down the holomorphic sections.

Before providing the detailed arguments, we summarize the holomorphic sections corresponding to various divisors of interest:
\be\label{SectionsX}
\begin{array}{|l||l| }\hline
\text{Holomorphic Section} & \hbox{Divisor class} \cr\hline\hline
Z_4 & H\cr\hline
Z_{1,2} & (H-E) + E =H \cr\hline
Z_3 & (H-D-E) + (D+E)= H \cr\hline
W_{1,2,3} & H-E\cr\hline
W_4 & 3H -D-2E \cr\hline
V_1 & ( 3H - D -2E) + E = 3 H - D - E \cr\hline
V_0 & H-D-E\cr\hline
\end{array}
\ee
The table also specifies whether a divisor is reducible.  For instance, $Z_3=0$ defines a pair of divisors in the classes $H-E$ and $E$ and hence is effectively in the class $H$.

We now turn to explain these identifications.

\begin{itemize}


\item \boxed{H:} \\
It is easy to see that $Z_4$ is a holomorphic section of $\mathcal{O}(H)$, i.e. the locus $Z_4=0$ is a holomorphic representative of $H$. It entirely misses the patch $Z_4=1$.  As such it misses the node of ${\cal{C}}$ and hence does not contain the conifold point $p_0$, two facts that are consistent with $H\cdot E=0$.


\item \boxed{H-D-E:} \\
We will show that  $V_0$ is a holomorphic section of $\mathcal{O}(H-D-E)$.
In an affine coordinate patch, which contains the node, the threefold takes the form
\be
V_0 (z_1z_2 + (z_1 + z_2)^3) = V_1 z_3 \,.
\ee
Since $V_0$ multiplies the cubic equation for ${\cal{C}}$, the divisor $V_0=0$ contains the full nodal curve.  This ensures that the divisor as a summand picks up a $-Q=-(D+E)$ from the blowup.
Furthermore, the corresponding divisor contains only the south pole of the $\mathbb{P}^1_V$, $[V_0,V_1]=[0,1]$, and thus misses the conifold point and will not intersect $E$. This is consistent with the intersection number $(H-D-E)\cdot E=0$.


\item \boxed{(H-E) + E:} \\
We will show that  $Z_1$ is a section of $\mathcal{O}(H)$, and that $Z_1 =0$ is a reducible divisor.
The equations (\ref{ZinW}) become
\be
\ba
0&= z_1 W_j  =  z_i W_1  \qquad \Rightarrow z_i =0 \quad \hbox{or} \quad W_1=0 \cr
0&= z_1W_4 = u W_1 \qquad \Rightarrow u=0\quad \hbox{or} \quad W_1=0\,.
\ea
\ee

In the patch $W_1=1$, this implies $u=z_i=0$ for all $i$ so that the condition $(z_1,z_2,z_3,u)\in [W_1,W_2,W_3,W_4]$ is vacuous.  This means that the entire exceptional divisor $E$ is contained here.  The second component of the solution is $W_1=0$, which contains $p_0$ and hence corresponds to $H-E$.  Note that $W_1=0$ parametrizes precisely the union of the two $\mathbb{P}^1$'s in $E$ so that the $W_1=0$ component of $Z_1=0$ intersects $E$ in $2\ell$, as we expect for a divisor in the class $H-E$.

Similar arguments hold for $Z_2=0$  and $Z_3=0$.


\item \boxed{H-E} \\
In order to show that  $W_1$ is a holomorphic section of $\mathcal{O}(H-E)$, note that
$W_1 z_i = z_1 W_i$ implies that either $z_1=0$ or $W_i=0$ for all $i$. The latter is clearly  not a solution, and thus, we are left with the solution space $z_1=0$. Not all solutions of $z_1=0$ are automatically solutions to $W_1=0$.  Rather, as we have seen for the discussion of the divisor $Z_1=0$, only the component which gives rise to $H-E$ is a solution of both $W_1=0=z_1$.


\item \boxed{3 H-D- 2 E} \\
To show that $W_4$ is a holomorphic section of  $\mathcal{O}(3H-D-2 E)$, note that $W_4=0$
implies
\be
W_1 W_2 = W_3 W_4 =0\qquad  \Rightarrow \qquad W_1 =0 \quad \hbox{or} \quad W_2=0 \,,
\ee
and thus intersects $E$ in two $\mathbb{P}^1$'s, consistent with the fact that $(3H - D- 2E )\cdot E =2 \ell$.
Furthermore, $W_4=0$ is equivalent to
 $v_1/v_0 =0$. The solution space $v_1=0$ (with $v_0 \not=0$) is given by setting the equation for the curve, ${\cal{C}}$, which is cubic in the coordinates $z_i$, to zero. Furthermore this divisor passes through the conifold point.


\item \boxed{(3 H-D- 2 E) + E} \\
The divisor $V_1=0$ has one component given by the solution space to
$W_4=0$, which we identified earlier with $3H- D- 2 E$.
The second component of this divisor has $z_1=z_2=z_3=v_1=0$, and thus, no constraint on the $W_i$ apart from (\ref{WWeq}), so that this component is precisely the divisor $E$.

\end{itemize}


\subsubsection{Topology of the Divisors}

A useful basis of divisors is $H$, $E$ and $H-D-E$. Their topology is
\be
\ba
H &\cong dP_3\cr
E &\cong \mathbb{P}^1 \times \mathbb{P}^1 \cr
H-D-E &\cong \mathbb{P}^2 \,.
\ea
\ee
By construction $E$ is the product of two $\mathbb{P}^1$'s as in (\ref{P1s}).
The divisor $H$ arises from a generic $\mathbb{P}^2$ inside $Z=\mathbb{P}^3$ which intersects the cubic curve ${\cal{C}}$ at three points away from the node.  After the blow-ups, then, $H$ corresponds to a $dP_3$.  Finally, $H-D-E$ originates from a $\mathbb{P}^2$ that contains the nodal curve ${\cal{C}}$.  The topology of this divisor is unaffected by the blow-ups.  Further details on the geometry of the divisors $Q$ and $D$ are summarized in Appendix \ref{app:Geo}.

This concludes our analysis of the three-fold $X$. From the data that we have extracted for $X$, we can now easily determine the full geometry of the final three-fold $\tilde{X}$ by following through the flop, as we do next.


\subsection{Embedding into $\mathbb{P}^3$: After the Flop}

\begin{figure}
\begin{center}
\epsfig{file=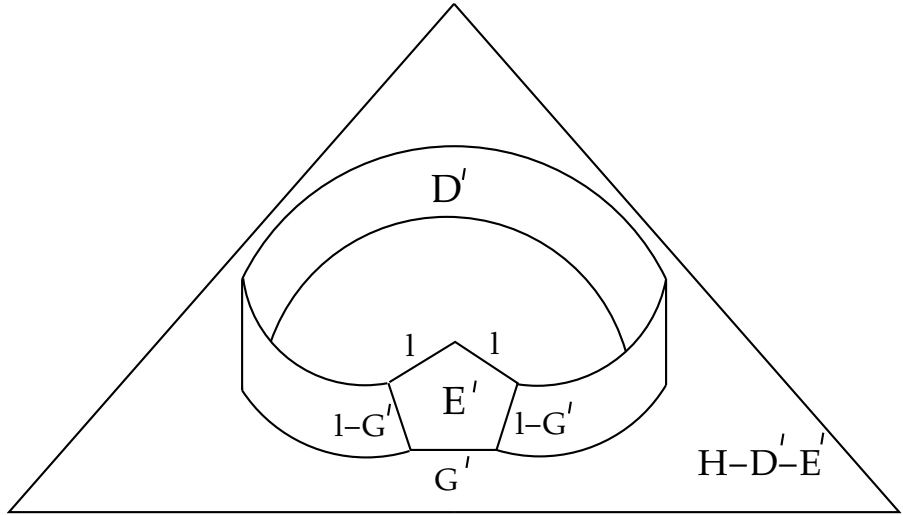,width=.7\textwidth}
\caption{Final  three-fold $\tilde{X}$}
\label{fig:GlobalThree}
\end{center}
\end{figure}

The curve $G$ is a $(-1,-1)$ curve because it is an exceptional $\mathbb{P}^1$.  This can also be seen directly by computing the normal bundle of $G$ inside $D$, which has degree $D\cdot G=-2$.  Because $G$ is $(-1,-1)$, we can flop it to obtain a new three-fold, $\tilde{X}$, depicted in figure \ref{fig:GlobalThree}.  The divisors $D$ and $E$ of $X$ carry over to new divisors $D'$ ad $E'$ in $\tilde{X}$.  The canonical class also follows simply from $K_X$ as
\be\label{KXtilde}
K_{\tilde{X}} = - 4 H + D' + 2 E' \,.
\ee
The resulting three-fold $\tilde{X}$ has  the desired property that the two curves $\ell-G'$ are distinct in $H_2(E',\mathbb{Z})$ but are nonetheless equivalent in $H_2(\tilde{X},\mathbb{Z})$ so that they satisfy the condition for existence of a suitable hypercharge flux \eqref{HyperConst}. 


\subsubsection{Topological Properties of $\tilde{X}$}
\label{subsec:Ints}

We now turn to a discussion of several important properties of $\tilde{X}$, including the topology of various divisors and the intersection tables for divisors and curves.  We start with a discussion of several interesting divisor classes.  The divisor $H$, which was a $dP_3$ before the flop, remains a $dP_3$ because it is unaffected by the flop.  From the viewpoint of $H-D-E=\mathbb{P}^2$, however, the flop corresponds to blowing up a point so that $H-D'-E'$ becomes a $dP_1$.  Similarly, from the viewpoint of $E=\mathbb{P}^1\times\mathbb{P}^1$, the flop effectively blows up a point so that $E'$ is simply $dP_2$.  The topology of the divisor $D'$ is analyzed in Appendix \ref{app:Geo} where it is shown to be the Hirzebruch surface $\mathbb{F}_4$.  We can now summarize these results as
\be\label{DivBasis}
\ba
H &\cong dP_3 \cr
E' &\cong dP_2\cr
D' &\cong\mathbb{F}_4 \cr
H-D'-E' &\cong dP_1 \,.
\ea
\ee

As a basis of $H_2(\tilde{X},\mathbb{Z})$, we take the curves $\ell_0$ and $\ell$ along with the flopped curve $G'$ as depicted in figure \ref{fig:GlobalThree}.  The intersection numbers of these curves with various divisors are presented in the following table

\begin{center}
\begin{tabular}{|r||r|r|r|r|}
\hline
${}$        & $H$ & $E'$ & $H-D'-E'$ & $D'$ \cr\hline\hline
$\ell_0$    & $1$ & $0$ & $+1$  & $0$ \cr\hline
$\ell$      & $0$ & $-1$ & $0$  & $1$\cr\hline
$G'$    & $0$ & $-1$ & $-1$ & $2$\cr\hline
$\ell-G'$   & $0$ & $0$ & $+1$ & $-1$\cr\hline
\end{tabular}
\end{center}

\noindent The intersections of the divisors with one another are as follows

\begin{center}
\begin{tabular}{|r||c|c|c|}
\hline
        &$H$    & $E'$  &   $H-D'-E'$   \cr\hline\hline
$H$     & $\ell_0   $& $0 $& $\ell_0 - 3l+ 3 G' $    \cr\hline
$E'$        & $0$   &$ - 2\ell+ G'$ & $G'$  \cr\hline
$H-D'-E'$   & $\ell_0 - 3\ell + 3 G'$ & $G'$ & $- 2 \ell_0+ 6 \ell - 5 G' $  \cr\hline
$D'$    & $3 \ell- 3 G'$ &  $2\ell - 2 G' $ & $3 \ell_0 - 9 \ell+ 7 G'$ \cr\hline
\end{tabular}
\end{center}

\noindent
It will in fact be useful to distinguish the two $\mathbb{P}^1$'s of $E'$ that are equivalent to $\ell$ inside $\tilde{X}$.  Denoting these by $\ell_1$ and $\ell_2$, we find that\footnote{In order to properly define this, we would require a refined notion of homology. For our purposes, it will suffice to notice that the divisor $D$ intersects $E$ in both $\mathbb{P}^1$s and thus, the same holds for $D'$ after the flop. We denote each of the $\mathbb{P}^1$s by $\ell_i -G'$.}
\be\label{NewEll}
{E'}^2 = G' - \ell_1 -\ell_2 \,,\qquad
D'.E' = (\ell_1 - G') + (\ell_2 - G') \,.
\ee

The non-vanishing triple intersection numbers are easily computed from the above data with the following results
\be
\ba
H^3 &= 1 \cr
E^{\prime\,3}&= 1 \cr
D^{\prime\,3}&= -6 \cr
D^{\prime\,2}H &= -3 \cr
D^{\prime\,2}E' &= -2 \,.
\ea
\ee

In section \ref{subsec:DivX} we studied various holomorphic
divisors and their corresponding sections on $X$.  Each of these carries over to a divisor or section after the flop.  We will abuse notation in what follows and continue to use the labels $Z_i,W_j,V_k$ of \eqref{SectionsX} for the corresponding holomorphic sections on $\tilde{X}$.


\subsubsection{Geometry of $S_{\rm GUT}$}
\label{subsubsec:GeomSGUT}

To build F-theory models of supersymmetric GUTs from elliptic fibrations over $\tilde{X}$, we will realize the $SU(5)$ gauge group on the divisor $E'$:
\be
S_{\rm GUT} = E' \cong dP_2 \,.
\ee
We now discuss the geometry of this divisor in some more detail.  As mentioned above, we know that $E'\cong dP_2$ because the flop blows up a point inside the divisor $E=\mathbb{P}^1\times\mathbb{P}^1$.  One basis for $H_2(E',\mathbb{Z})$ is given by the curve classes $\ell_1,\ell_2$ inherited from $E$ along with the exceptional class $G'$.  In this basis, the intersection form inside $E'$ is specified by
\be 
\ell_1^2=\ell_2^2=\ell_i\cdot G'=0\,,
\qquad \ell_1\cdot\ell_2=1\,,
\qquad G^{\prime\,2}=-1\,.
\ee
In what follows, however, we prefer to use the standard basis for $dP_2$ consisting of the hyperplane class, $\tilde{h}$, and the two exceptional curves, $\tilde{e}_i$ and $\tilde{e}_j$
\be 
H_2(E',\mathbb{Z}) = \langle\tilde{h},\tilde{e}_1,\tilde{e}_2\rangle\,.
\ee
From the intersection form
\be
\tilde{h}^2=1\,,\qquad \tilde{e}_i\cdot \tilde{e}_j=-\delta_{ij}\,,
\ee
it is easy to obtain the standard relation of these classes to $\ell_1$, $\ell_2$, and $G'$,
\be\label{CurveIdent}
\ba
\ell_1 &= \tilde{h}-\tilde{e}_1 \\
\ell_2 &= \tilde{h}-\tilde{e}_2 \\
G' &= \tilde{h}-\tilde{e}_1-\tilde{e}_2\,.
\ea
\ee
Using \eqref{NewEll}, we can now determine the class inside $H_2(E',\mathbb{Z})$ of the intersection of any divisor in $\tilde{X}$ with $E'$




\begin{center}
\begin{tabular}{|r||c|}
\hline
& $E'$  \cr\hline\hline
$H$ & $0$ \cr \hline
$D'$ & $\tilde{e}_1 + \tilde{e}_2$ \cr\hline
$E'$ & $-\tilde{h}$ \cr\hline
$(H- D' -E')$&  $\tilde{h}- \tilde{e}_1 - \tilde{e}_2 $\cr\hline
\end{tabular}
\end{center}




Note that $S_{GUT}=dP_2$ is non-contractible in ${\tilde X}.$ However, it is possible
to get mass hierarchy $M_{GUT}\ll M_{pl}.$ Indeed, K\"ahler form $J$ on ${\tilde X}$ is given by
$$J=mH-aE'-bD'$$
with
$$m>0,\quad a>0,\quad b>0,\quad a>2b,$$
where $a\ge 2b$ ensures that the curve $G'$ has non-negative volume.  
Let us compute the volume of ${\tilde X}$:
$$Vol_{\tilde X}={1\over 6}\int_{\tilde X}J^3={1\over 6}\bigl(m^3-a^3+6b^3-9mb^2+6ab^2\bigr)$$
and the volume of $S_{GUT}$:
$$Vol_{S_{GUT}}=\half \int_{S_{GUT}}J^2=\half(a^2-2b^2)>0.$$
In the limit $m\gg a$ and $m\gg b$ one can get $M_{GUT}\ll M_{pl}.$
In this paper we did not study moduli stabilization, so it remains an open
question if $J$ with $m\gg a$ and $m\gg b$ naturally arises via stabilization. 


\subsection{Summary of Three-fold $\tilde{X}$}
\label{subsec:SummaryBase}

Let us summarize the geometry of the compact three-fold $\tilde{X}$ depicted in figure \ref{fig:GlobalThree}.
It has the following divisor classes
\be
\ba
S_{\rm GUT}= E' &: \qquad dP_2\cr
H &: \qquad dP_3 \cr
H-D'-E' &: \qquad dP_1 \cr
D' &: \qquad \phantom{d}\mathbb{F}_4 \,.
\ea
\ee
Its canonical class is
\be
K_{\tilde{X}} = - 4 H + D' + 2 E' \,.
\ee
The curves and triple intersections were determined in subsection \ref{subsec:Ints}.
We will take $E'$ to be the divisor on which the GUT gauge group is localized. The classes in $H_2 (E',\mathbb{Z})$ are identified with the curves in the three-fold as in (\ref{CurveIdent}).
The exceptional classes $\tilde{e}_1$ and $\tilde{e}_2$ inside $E'$ are by construction homologous in $\tilde{X}$ and are thus ideal candidates to use for the hypercharge flux
\be
[F_Y] =\tilde{e}_1 - \tilde{e}_2 \,.
\ee
It is in fact easy to see the explicit 3-chain that connects $\tilde{e}_1$ and $\tilde{e}_2$ inside $\tilde{X}$.  This is because each of these curves is homologous to the fiber class of $D'=\mathbb{F}_4$, which we recall is a $\mathbb{P}^1$ fibration over a base $\mathbb{P}^1$.  The 3-chain is constructed from the fiber class along with the obvious path inside the base $\mathbb{P}^1$.

Note, that the resulting three-fold is not Fano, but almost Fano, as the anti-canonical bundle is almost ample. In particular, the curve $G'$ satisfies 
\be
- K_X \cdot G' =0 \,.
\ee
This completes the discussion of the geometry of the three-fold base $\tilde{X}$.


\section{$SU(5)$ GUTs from Elliptic Fibrations over $\tilde{X}$}
\label{sec:SU5}

In this section, we turn to the study of elliptic fibrations over the three-fold base $\tilde{X}$.  We first review the global constraints of section \ref{sec:Global} in the context of this specific example.  We then proceed to describe the general structure of matter curves and demonstrate, using the results of \cite{Donagi:2009ra}, that it is possible to engineer 3 generations of chiral MSSM matter.  


\subsection{Global Constraints}

Let us first recall the global constraints described in section \ref{sec:Global}.  The 7-brane tadpole is taken care of by the Calabi-Yau nature of the four-fold, while the hypercharge flux condition has been built in by our construction of $\tilde{X}$.  Indeed, $S_{\rm GUT}$ is a $dP_2$ surface with exceptional classes $\tilde{e}_1$ and $\tilde{e}_2$ that are homologous in $\tilde{X}$ so that we can adopt the standard procedure of breaking the GUT group with a hypercharge flux $[F_Y]=\tilde{e}_1-\tilde{e}_2$ \cite{Beasley:2008kw,Donagi:2008kj}.

Recall, however, that the quantization condition for $G$-fluxes,
\begin{equation}[G_4]-\frac{c_2(Y_4)}{2}\in H^4(Y_4,\mathbb{Z})\,,
\end{equation}
gave us a nontrivial condition on the base manifold $\tilde{X}$.  In particular, to avoid Lorentz-violating $G$-fluxes which have all four indices along $\tilde{X}$, we needed
\begin{equation}c_2(Y_4)|_{\tilde{X}}=c_2(\tilde{X})-c_1(\tilde{X})^2\,,
\end{equation}
to be an even class.  For this, we can use the computation of $c_2(\tilde{X})$ in Appendix \ref{app:Xtilde}
along with the intersection tables contained therein to directly compute
\begin{equation}c_2(\tilde{X})-c_1(\tilde{X})^2=2\left(\ell-G'-2\ell_0\right)\,,
\end{equation}
which is a manifestly even class.

Finally, we should address the D3 tadpole (\ref{D3tad}).  For this, we recall that the Euler character of any smooth elliptic fibration over $\tilde{X}$ can be computed using the formula (\ref{D3tadsimp})
\begin{equation}
\chi = 12\int{\tilde{X}}c_1(c_2+30c_1^2)\,.
\end{equation}
For our specific $\tilde{X}$, this computation is easily performed, with the result
\begin{equation}\frac{\chi}{24} = 582\,.
\end{equation}
We expect that this accurately captures the induced D3 brane charge of our fibrations to follow.  That it is both integral and positive means that it can in principle be cancelled by adding some combination of D3 branes or $G$-fluxes.


\subsection{Matter Curves}

As described in section \ref{subsec:EllipticCY}, the structure of matter curves and their intersections is determined by the choice of five sections which we denoted $h,H,q,f_3,g_5$.  Using the result (\ref{KXtilde}) for the canonical class of $\tilde{X}$, the bundles of which these are sections of are as follows
\begin{equation}\begin{array}{c|c}
\text{Section} & \text{Bundle} \\\hline
h & 4H-D'-2E' \\
H & 8H-2D'-5E' \\
q & 12H -3D' - 8E' \\
f_3 & 16H-4D'-11E' \\
g_5 & 24H - 6D' - 17E'
\end{array}\end{equation}

Because the $SO(10)$ matter curve, $\Sigma_{10}$, is described by the intersection of $h=0$ with $S_{\rm GUT}$, we see that the class of $\Sigma_{10}$ inside $dP_2$ is
\begin{equation}\Sigma_{10} = 2\tilde{h}-\tilde{e}_1-\tilde{e}_2\,,
\end{equation}
which, when irreducible, has the topology of a $\mathbb{P}^1$.
The $SU(6)$ matter curve, $\Sigma_5$, is described by the intersection of $P=0$ with $S_{\rm GUT}$ where we recall that
\begin{equation}
P=-3Hq^2-3hf_3q+2g_5h^2 \,.
\end{equation}
We see that $P$ is a section of ${\cal{O}}(32H-8D'-21E')$,
\begin{equation}
P\in\Gamma({\cal{O}}(32H-8D'-21E'))\,,
\end{equation}
so that the class of $\Sigma_5$ inside $S_{\rm GUT}$ is
\begin{equation}\Sigma_5 = 21\tilde{h} - 8(\tilde{e}_1+\tilde{e}_2)\,.
\end{equation}


\subsection{Getting Three Generations}
\label{sec:ThreeGen}

In this subsection, we demonstrate that, under suitable conditions on the fibration, it is possible to turn on  $G$-fluxes to obtain a model with 3 chiral generations of MSSM matter.  This analysis relies heavily on the recent results of \cite{Donagi:2009ra}, where it was demonstrated that one can use the fundamental spectral cover $\bar C$ of $S_{\rm GUT}$
to determine allowed $G$-fluxes in F-theory.
Let  $\bar C$ be a 5-sheet cover of $S_{\rm GUT}$  defined by the
following equation in an auxiliary three-fold $\mathcal{M}_3$--the total space of the projective bundle  $\mathbb{P}\left(\mathcal{O}_{S_{\rm GUT}}\oplus K_{S_{\rm GUT}}\right)$ over $S_{\rm GUT}$:
\be
b_0u_2^5+b_2u_1^2u_2^3+\ldots b_5u_1^5=0 \,.
\ee
Here $u_1,u_2$ are homogenous coordinates on the $\mathbb{P}^1$ fiber of  $\mathcal{M}_3,$ so that
$u_2$($u_1$) is a section of $K_{S_{\rm GUT}}$($\mathcal{O}_{S_{\rm GUT}}$). Note that $b_0,\ldots,b_5$
are related with $h,H,q,f_3,g_5$ as in (\ref{Tate}).

It was shown in \cite{Donagi:2009ra}  that the independent $G$-fluxes allowed in F-theory are
in one to one correspondence with integral classes $\gamma$ in $H^{(1,1)}(\bar C)$ such that
\be \label{restriction} 
p_C: \ {\bar C} \rightarrow S_{\rm GUT}\,,\qquad p_{C*}\gamma=0 \,.
\ee
Further,  for generic $b_0,\ldots,b_5$ there is only one independent
class $\gamma_u$, the  so-called universal class.
To obtain three generations one should choose $b_0,\ldots b_5$ in a specific way
such that there is at least one primitive class in $H^{1,1}(\bar{C})$ with the property
(\ref{restriction}). According to \cite{Donagi:2009ra}, in order to construct such a primitive class
one should first find a curve $\alpha_0$ in $S_{\rm GUT}$ which intersects $\Sigma_{10}$
and then construct a curve $\alpha$ in $\bar C$ which does not intersect $\Sigma_{10}$  and covers
$\alpha_0$ precisely once. Then, the primitive class is given by
\be
\gamma_{prim}=5\alpha-p_C^*p_{C*}\alpha \,,\ee
and
the number of chiral generations is
\be
N_{chir}=-\half \int_{\Sigma_{10}}\gamma_u -n\int_{\Sigma_{10}}\gamma_{prim}\,.
\ee
The computation of integrals in the above formula reduce to the
intersection of classes in $S_{\rm GUT}$:
\be
\ba
\int_{\Sigma_{10}}\gamma_u&=-(6c_1(S_{\rm GUT})-t)\cdot(c_1(S_{\rm GUT})-t) \cr
\int_{\Sigma_{10}}\gamma_{prim}&=\alpha_0\cdot\Sigma_{10} \,,
\ea
\ee
where $t=-c_1(N_{S_{\rm GUT}\vert X_3})$.

In our case we may choose $\alpha_0$ in the class $h$ in
$S_{\rm GUT}=dP_2,$ so that $\alpha_0$ arises as the intersection of the
divisors $H-E'$ and $E'$ in ${\tilde X}$. Then, we have $\alpha_0
\cdot \Sigma_{10}=2.$ Furthermore, in our construction 
\be
c_1(S_{\rm GUT})=3\tilde{h}-\tilde{e}_1-\tilde{e}_2,\quad t=\tilde{h}\,,
\ee
so that $\int_{\Sigma_{10}}\gamma_u=-22$ and we must choose $n=4$ to get three generations.

The non-generic choice of $b_0$ in our case is very similar to that of
\cite{Donagi:2009ra}, but we prefer to work with the global description in terms of
sections on $X_3$.
 More concretely, in the auxiliary four-fold $\mathcal{M}_4 $, the
total space of the projective bundle
$\mathbb{P}\left(\mathcal{O}_{X_3}\oplus \left(K_{X_3}\otimes S_{\rm GUT}\right)\right)$ over $X_3$, 
the curve $\alpha_0$ is given by
\be
W_3=0\,,\qquad z_{S_{\rm GUT}}=0\,,\qquad u_2=0 \,.
\ee
The curve $\alpha$ is constructed as in \cite{Donagi:2009ra} as 
\be
W_3=0\,,\qquad z_{S_{\rm GUT}}=0\,,\qquad u_1={\cal P}u_2 \,,
\ee
where ${\cal P}$ is a section of $K_{X_3}^{-1}\otimes S_{\rm GUT}^{-1}.$ Then, the condition
that $\alpha \in \bar C$ gives a constraint on $b_0$:
\be
b_0=-\Bigl(b_2{\cal P}^2+\ldots b_5{\cal P}^5\Bigr)\vert_{W_3\rightarrow 0}+O(W_3)\,.
\ee
Using the relation between $b_0,\ldots,b_5$ and $h,q,H,f_3,g_5$ as in (\ref{Tate})
we recast this constraint as
\be
g_5=-18 {\cal P}^2\Bigl(f_3+4q{\cal P}-12H{\cal P}^2+48h{\cal P}^3\Bigr)\vert_{W_3\rightarrow 0}+O(W_3) \,.
\ee



\section{Getting Flavor Hierarchies}
\label{sec:Realistic}

Though it is possible to construct examples with three generations of $\mathbf{10}_M$ and $\mathbf{\overline{5}}_M$ matter fields, phenomenologically successful models require significantly more structure.  As detailed in section \ref{sec:Loc}, this involves putting $\mathbf{\overline{5}}_M$, $H_u$, and $H_d$ on distinct $\mathbf{5}$ matter curves with the right intersection properties.  To build realistic models, then, it is first necessary to find a geometric setup in which the matter curve $\Sigma_5$ is reducible and splits into components with suitable candidates for the $\mathbf{\overline{5}}_M$, $H_u$, and $H_d$ matter curves.  Given a family of four-folds with this property, the next step would be to demonstrate the existence of suitable $G$-fluxes that engineer the MSSM matter fields in the ``correct" place.  In this section, we turn our attention to the geometric conditions and demonstrate that it is not difficult to satisfy them.  We hope to address the issue of $G$-fluxes in future work.

\subsection{Geometric Conditions}

In addition to requiring $H_u$, $H_d$, and the MSSM $\mathbf{\overline{5}}_M$ matter fields to localize on distinct matter curves, studies of flavor structure in the context of local models suggests that the up-type and down-type Yukawas should each originate from a \emph{unique} point of singularity enhancement (either $E_6$ or $SO(12)$) where appropriate matter curves meet.  In general, however, our four-folds have several points in $S_{\rm GUT}$ that exhibit $E_6$ and $SO(12)$ singularities in the fiber.  For instance, recall that we obtain $E_6$ singularities at points inside $S_{\rm GUT}=E'$ where the holomorphic sections $h$ and $H$ simultaneously vanish.  Because $h$ is a section of ${\cal{O}}(4H-D'-2E')$, $h=0$ intersects $S_{\rm GUT}$ in a curve in the class $2\tilde{h}-\tilde{e}_1-\tilde{e}_2$.  On the other hand, $H$ is a section of ${\cal{O}}(8H-2D'-5E')$ so that $H=0$ intersects $S_{\rm GUT}$ in the class $5\tilde{h}-2(\tilde{e}_1+\tilde{e}_2)$.  These two curves have intersection number 6 inside $S_{\rm GUT}$ so that there are generically 6 points with an $E_6$ singularity in the fiber.

Similarly, $SO(12)$ singularities arise at points inside $S_{\rm GUT}$ where the holomorphic sections $h$ and $q$ simultaneously vanish.  The divisor $q=0$ is in the class $12H-3D'-8E'$ so that it intersects $S_{\rm GUT}$ in the class $8\tilde{h}-3(\tilde{e}_1+\tilde{e}_2)$.  This means that there are generically 10 points inside $S_{\rm GUT}$ with an $SO(12)$ singularity in the fiber.

Our goal is now clear.  We seek an elliptic fibration over $\tilde{X}$ in which $\Sigma_5$ factorizes into several irreducible components so that these $SO(12)$ and $E_6$ singular points are sufficiently spread out.  Among these components we seek 3, denoted $\Sigma_{H_u}$, $\Sigma_{H_d}$, and $\Sigma_{\overline{5}}$, such that $\Sigma_{H_u}$ meets $\Sigma_{10}$ in exactly 1 $E_6$ point and $\Sigma_{H_d}$ meets $\Sigma_{\overline{5}}$ and $\Sigma_{10}$ in exactly 1 $SO(12)$ point.  To achieve relatively minimal mixing in the quark sector, it is further preferable for these two singular points to be fairly close to one another inside $S_{\rm GUT}$.  In fact, one could imagine merging these two points into a single point that exhibits a singularity of higher rank, such as $E_7$ or $E_8$, in the fiber \cite{Bouchard:2009bu}.

A further condition that we find appealing, though not necessary, is to look for four-folds in which all irreducible components of $\Sigma_5$ are curves of genus 0.  Such a large splitting of $\Sigma_5$ is more likely to yield a significant spread of $E_6$ and $SO(12)$ points among various components and also has the nice advantage that each component will house a purely chiral spectrum.  This means that, with the right $G$-fluxes, it is feasible that one could engineer precisely the chiral matter content of the MSSM with nothing else, not even extra vector-like pairs{\footnote{Of course, one usually does not worry about vector-like pairs unless there is an additional symmetry that protects them.  Nevertheless, we feel that a model without such pairs would be nice to attain.}}.



\subsection{Candidate Geometries for F-theory GUTs}

We therefore seek elliptic fibrations over $\tilde{X}$ in which the holomorphic section, $P$, that determines $\Sigma_5$ factorizes into a product of holomorphic sections
\begin{equation}
P = \prod_i P_i \,.
\label{Pfactorization}\end{equation}
For generic $P_i$, the divisor defined by $P_i=0$ will meet $S_{\rm GUT}$ in an irreducible curve class.  However,  it is possible that such a matter curve is further reducible inside $S_{\rm GUT}$.  Several examples of this are discussed in detail in Appendix \ref{app:curveexamples}.  For clarity, we will use the term ``factor" in what follows for the $P_i$, which may or may not be irreducible inside $S_{\rm GUT}$.  We will reserve the term ``component" for an irreducible component of $P$.

This distinction is important because the reducibility of at least one factor $P_i=0$ inside $S_{\rm GUT}$ is crucial for model building with elliptic fibrations over $\tilde{X}$, as we now explain.  Any divisor inside $\tilde{X}$ intersects $S_{\rm GUT}$ in a class that is symmetric in $\tilde{e}_1$ and $\tilde{e}_2$ so that, in particular, the GUT-breaking hypercharge flux, $[F_Y]\sim \tilde{e}_1-\tilde{e}_2$, restricts trivially to the matter curve originating from each factor, $P_i$.  The distinguishing feature of a Higgs matter curve, though, is a nontrivial hypercharge flux that lifts the unwanted triplets.  This means that to obtain matter curves for Higgs fields, we must require at least one of the factors $P_i=0$ to be reducible inside $S_{\rm GUT}$ into components with nontrivial $F_Y$.

In fact, the absence of exotics requires that $H_u$ and $H_d$ come from two distinct components of the same factor.  That the hypercharge flux restricts trivially to a given factor means that there is no net chirality in the doublet spectrum there{\footnote{To simplify the discussion, we assume that the $\mathbf{\overline{5}}_M$ matter fields are localized on a different factor so that the net chirality of $\mathbf{5}$'s on the Higgs factor is zero.  This assumption is ultimately not necessary, though.}}.  This means that if $H_u$ is engineered on one component of the factor, another component will necessarily contain a second light doublet with opposite quantum numbers.  If this extra doublet is not $H_d$ then it represents an additional unwanted exotic.

We now turn to the issue of realizing a particular factorization \eqref{Pfactorization} inside an elliptically fibered four-fold.  What makes this slightly nontrivial is that we do not specify $P$ directly when constructing the fibration but rather the sections $h$, $H$, $q$, $f_3$, and $g_5$, in terms of which $P$ is given by \eqref{PQdef}
\begin{equation}
P=-3Hq^2-3hf_3q+2g_5h^2\,.
\end{equation}
In Appendix \ref{app:Fibs}, we discuss a general method for constructing four-folds for a given choice of factors, $P_i$, before going on to discuss several explicit examples.  In all of these examples, $h$ is chosen so that the $\mathbf{10}$ matter curve, $\Sigma_{10}$, is an irreducible element of the class $2\tilde{h}-\tilde{e}_1-\tilde{e}_2$
\begin{equation}
\Sigma_{10} = 2\tilde{h}-\tilde{e}_1-\tilde{e}_2\,.
\end{equation}


\subsubsection{The ``Maximal" Factorization}

The most natural factorization to consider is the ``maximal" one, in which the $P_i$ all reduce inside $S_{\rm GUT}$ to linear polynomials in the $W_j$.  This is the example that we discuss in appendix  \ref{app:Ex1}.  As described there, it corresponds to a splitting of $\Sigma_5$ into factors according to
\begin{equation}\Sigma_5=21\tilde{h}-8(\tilde{e}_1+\tilde{e}_2)\rightarrow 8\times (2\tilde{h}-\tilde{e}_1-\tilde{e}_2) + 5\times \tilde{h}\,.\end{equation}
The factors of $\Sigma_5$ naturally split into three groups, which we denote as in the following table
\begin{equation}\begin{array}{c|c}\text{Matter Curve} & \text{Class in }dP_2 \\ \hline
\Sigma_{10} & 2\tilde{h}-\tilde{e}_1-\tilde{e}_2 \\
\Sigma_{5,a}\,\,(a=1,\ldots,5) & 2\tilde{h}-\tilde{e}_1-\tilde{e}_2 \\
\Sigma_{5,A}\,\,(A=1,\ldots,3) & 2\tilde{h}-\tilde{e}_1-\tilde{e}_2 \\
\Sigma_{5,\mu}\,\,(\mu=1,\ldots,5) & \tilde{h}
\end{array}\end{equation}
We also include the $\mathbf{10}$ matter curve in this table for completeness.  In constructing the fibration, the intersections of various factors of $\Sigma_5$ with $\Sigma_{10}$ are fixed as in the following table
\begin{equation}\begin{array}{c|c}\text{Intersecting Curves} & \text{Singularity Types} \\ \hline
\Sigma_{10}\cap \Sigma_{5,a}\cap \Sigma_{5,\mu}\,\,\,(\mu=a) & SO(12) \\
\Sigma_{10}\cap \Sigma_{5,A} & E_6
\end{array}\end{equation}
Each of these intersections occurs with multiplicity 2 so it is easy to verify that there are generically 6 $E_6$ points and 10 $SO(12)$ points.  An important feature of this example is that no single factor of $\Sigma_5$ contains both an $E_6$ point and an $SO(12)$ point.  Because the $H_u$ matter curve must contain an $E_6$ point and the $H_d$ matter curve must contain an $SO(12)$ point, there is no single candidate ``Higgs factor" from which both may be obtained as components.  As described above, this means that we will be forced to introduce extra exotics beyond the matter content of the MSSM.


\subsubsection{A Geometry with Hierarchical Yukawas}

The problem with the previous example was that $\Sigma_5$ was ``too factorized" in the sense that no single factor contained both $SO(12)$ and $E_6$ points.  A possible remedy for this is to consider a slightly less factorized form in which one of the factors corresponds to a (possibly degenerate) curve of genus 1 inside $S_{\rm GUT}$.  If this curve contains both $SO(12)$ and $E_6$ points then it can be a good candidate Higgs curve if it is indeed degenerate, thereby splitting into distinct components for $H_u$ and $H_d$ inside $S_{\rm GUT}$.

We consider two different Ans\"atze for $P$ of this type in Appendix \ref{app:Ex2} and \ref{app:Ex3} which lead to two families of elliptically fibered four-folds.  We focus here on the example of section \ref{app:Ex2}, in which $\Sigma_5$ is split according to
\begin{equation}\Sigma_5=21\tilde{h}-8(\tilde{e}_1+\tilde{e}_2)\rightarrow (3\tilde{h}-\tilde{e}_1-\tilde{e}_2)+7\times (2\tilde{h}-\tilde{e}_1-\tilde{e}_2)+4\times \tilde{h}\,.
\end{equation}
The factors of $\Sigma_5$ naturally divide into five groups which we denote as in the table below
\begin{equation}\begin{array}{c|c}\text{Matter Curve} & \text{Class in $dP_2$} \\ \hline
\Sigma_{10} & 2\tilde{h}-\tilde{e}_1-\tilde{e}_2 \\
\Sigma_{5,a}\,\,\, (a=1,\ldots,4) & 2\tilde{h}-\tilde{e}_1-\tilde{e}_2 \\
\Sigma_{5,A}\,\,\, (A=1,2) & 2\tilde{h}-\tilde{e}_1-\tilde{e}_2 \\
\Sigma_{5,\mu}\,\,\, (\mu=1,\ldots,4) & \tilde{h} \\
\tilde{\Sigma}_{5,\mathbb{P}^1} & 2\tilde{h}-\tilde{e}_1-\tilde{e}_2 \\
\tilde{\Sigma}_{5,T^2} & 3\tilde{h}-\tilde{e}_1-\tilde{e}_2
\end{array}\end{equation}
As before, we include the $\mathbf{10}$ matter curve, $\Sigma_{10}$, for completeness.  The intersections of $\Sigma_{10}$ with various components of $\Sigma_5$ are as follows
\begin{equation}\begin{array}{c|c}\text{Intersecting Curves} & \text{Singularity Types} \\ \hline
\Sigma_{10}\cap \Sigma_{5,a}\cap \Sigma_{5,\mu}\,\,\,(\mu=a) & SO(12) \\
\Sigma_{10}\cap \Sigma_{5,A} & E_6 \\
\Sigma_{10}\cap \tilde{\Sigma}_{5,\mathbb{P}^1} \cap \tilde{\Sigma}_{5,T^2} & SO(12) \\
\Sigma_{10}\cap \tilde{\Sigma}_{5,T^2} & E_6
\end{array}\end{equation}
Once again, it is easy to verify that there are generically 6 $E_6$ points and 10 $SO(12)$ points.  Note that if we identify $\Sigma_{5,\mathbb{P}^1}$ as the $\mathbf{\overline{5}}_M$ matter curve and $\tilde{\Sigma}_{5,T^2}$ as a Higgs matter curve then we get precisely the Yukawa couplings that are needed for the MSSM.  Further, we demonstrate in Appendix \ref{app:Ex2} that it is possible to choose $\tilde{\Sigma}_{5,T^2}$ to be a reducible curve inside $S_{\rm GUT}$ that splits according to
\begin{equation}\tilde{\Sigma}_{5,T^2} = 3\tilde{h}-\tilde{e}_1-\tilde{e}_2 \rightarrow (2\tilde{h}-\tilde{e}_1) + (\tilde{h}-\tilde{e}_2) \,.
\end{equation}
For this choice, one can keep track of the intersections of each component.  The results can be summarized as follows
\begin{equation}\begin{array}{c|c|c}\text{Intersecting Curves} & \text{Singularity Type} \\ \hline
\Sigma_{10} \cap \tilde{\Sigma}_{5,\mathbb{P}^1}\cap \tilde{\Sigma}_{5,T^2}^{(2\tilde{h}-\tilde{e}_1)} & SO(12) \\
\Sigma_{10} \cap \tilde{\Sigma}_{5,T^2}^{(2\tilde{h}-\tilde{e}_1)} & E_6 \\
\Sigma_{10} \cap \tilde{\Sigma}_{5,T^2}^{(\tilde{h}-\tilde{e}_2)} & E_6
\end{array}\end{equation}
Moreover, each of the intersections in the above table corresponds to a single point{\footnote{In general, we expect that $\Sigma_{10}$ (in the class $2\tilde{h}-\tilde{e}_1-\tilde{e}_2$) meets $\tilde{\Sigma}_{5,T^2}$ (in the class $3\tilde{h}-\tilde{e}_1-\tilde{e}_2$) in 4 points.  In that sense, the specific choice of matter curves in Appendix \ref{app:Ex2} is nongeneric as one intersection point, namely the $SO(12)$ one, occurs with degree 2.}}.  Given this structure, we can satisfy all of the requirements set forth in section \ref{sec:Loc} provided we identify
\begin{equation}\begin{array}{c|c}\text{Curve} & \text{Matter to be Engineered} \\ \hline
\Sigma_{10} & \mathbf{10}_M \\
\tilde{\Sigma}_{5,\mathbb{P}^1} & \mathbf{\overline{5}}_M \\
\tilde{\Sigma}_{5,T^2}^{(2\tilde{h}-\tilde{e}_1)} & H_d \\
\tilde{\Sigma}_{5,T^2}^{(\tilde{h}-\tilde{e}_2)} & H_u
\end{array}\end{equation}
In particular, the up-type Yukawa coupling arises from a single $E_6$ point and the down-type Yukawa coupling from a single $SO(12)$ point{\footnote{Though the $H_d$ matter curve meets the $\mathbf{10}_M$ one at a second $E_6$ point, $SU(3)\times SU(2)\times U(1)_Y$ invariance prohibits any couplings involving only massless modes from arising there}}.

In summary, this example contains candidate matter curves for all MSSM matter fields which exhibit the right intersection properties to give rise to Yukawa couplings with natural hierarchies.  It remains to demonstrate that suitable $G$-fluxes can be introduced in order to localize matter fields on the specific curves we have assigned to them.


\section{Concluding Remarks}

In this paper, we have studied F-theory compactifications on elliptically fibered four-folds that give rise to $SU(5)$ supersymmetric GUTs.  The three-fold $\tilde{X}$ that we constructed to serve as the base of these fibrations is remarkably simple, having only two divisor classes in addition to the contractible $dP_2$ on which the GUT gauge group is localized.  Because of this simplicity, we expect that four-folds constructed from $\tilde{X}$ will provide a fruitful setting for exploring the properties of compact supersymmetric GUT models in F-theory.  

For generic four-folds  that are built from $\tilde{X}$ and realize $SU(5)$ GUTs, we described global consistency conditions, studied the matter curves and their intersections, and demonstrated that, with a suitable tuning, it is possible to engineer precisely 3 chiral generations of MSSM matter.
To improve the phenomenology of these compactifications, though, it is necessary to refine them in order to accommodate some of the intricate geometric structures that have been suggested by the study of local models.  In the present work, we have considered some simple conditions that give rise to flavor hierarchies in the Yukawa matrices and a sufficiently long lifetime for the proton.  To build truly realistic compact models, however, several issues remain to be addressed.


Firstly, it is necessary to demonstrate that the realization of geometric ingredients from local models along the lines of section \ref{sec:Realistic} can be combined with the existence of $G$-fluxes that enable one not only to engineer 3 generations of the MSSM, but to ensure that the various matter fields localize on the ``correct" curves.  In addition, more intricate phenomenological questions beyond those considered here should be addressed.  For instance, we have paid virtually no attention to the pairwise intersection of $\mathbf{5}$ matter curves in $S_{\rm GUT}$ at points with an $SU(7)$ degeneration of the fiber.  The interactions that originate from such points serve a number of useful purposes, from generating Dirac neutrino masses \cite{Beasley:2008kw,Bouchard:2009bu,Randall:2009dw} to providing a simple implementation of gauge mediation \cite{Marsano:2008jq}.  Along these lines, it would also be interesting to study various supersymmetry-breaking mechanisms, such as the one proposed in \cite{Heckman:2008es}, in a compact example since this would avoid many of the intrinsic assumptions that have to be made when discussing this issue in a local setting.

Finally, variations on the construction  in section \ref{sec:ThreeFold}, by choosing embeddings of the cubic curve into three-folds other than $\mathbb{P}^3$, should give rise to  a large class of almost Fano three-folds that would be interesting to study in the same way as $\tilde{X}$.







\section*{Acknowledgements}

We are especially grateful to Melissa Liu for helping us to understand various aspects of the geometry of $\tilde{X}$.
We would also like to thank V.~Braun, A.~Grassi, S.~Gukov, K.~Matsuki, S.~Sethi, and T.~Weigand, for valuable discussions.
We thank C. Cordova for reminding us to update the arxiv version of the paper to include the correct statement already present in the JHEP version, that the three-fold is almost Fano (rather than Fano) . 
The work of JM and SSN was supported by John A. McCone Postdoctoral Fellowships.
The work of NS was supported in part by the DOE-grant DE-FG03-92-ER40701.
JM is grateful to the Institute of Mathematical Sciences in Chennai, the Chennai Mathematical Institute, the organizers of the 2008 Indian Strings Meeting, and the Enrico Fermi Institute at the University of Chicago for their hospitality at various stages of this work.  SSN thanks the KITP, Santa Barbara, and ENS, Paris, for hospitality during the course of this work.


\newpage

\appendix


\begin{figure}[h]
\begin{center}
\includegraphics[scale=1]{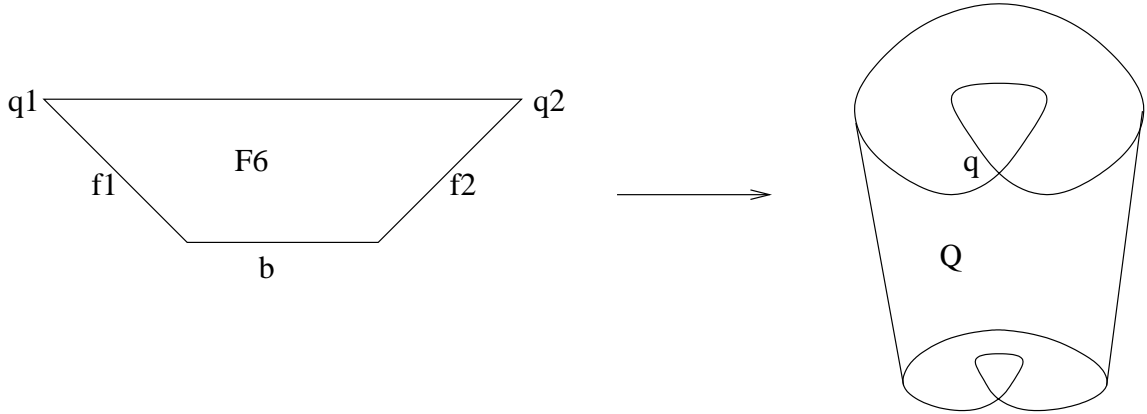}
\caption{Resolution of singularity  $\pi_Q: \mathbb{F}_6 \to Q$.}
\label{fig:F6}
\end{center}
\end{figure}


\section{Topology of various divisors}
\label{app:Geo}

\subsection{Topology of $Q$ and $D$}

In this appendix we discuss the geometries of various divisors in $X$ and $\tilde{X}$. 
Let $i:{\cal C}\to \mathbb{P}^3$ be the inclusion
of the nodal curve defined by 
\be
\begin{split}
Z_4 Z_1 Z_2 + (Z_1+ Z_2)^3 & =0,\\
Z_3 &=0.
\end{split}
\ee
The nodal curve is the zero locus of a cubic equation
and a linear equation, so the normal bundle of ${\cal C}$
in $\mathbb{P}^3$ is given by 
\be
N_{{\cal C}/\mathbb{P}^3} = L ^3 \oplus L,\quad L=i^*{\cal O}_{\mathbb{P}^3}(1).
\ee
So the exceptional divisor $Q$ is the following $\mathbb{P}^1$-bundle over $\mathcal{C}$: 
\be
Q \cong \mathbb{P}(N_{{\cal C}}) =\mathbb{P}(L^3\oplus L)\cong \mathbb{P}(L^2\oplus \mathcal{O}_{\mathcal{C}}).
\ee
Let $\pi_{\mathcal{C}}: \mathbb{P}^1 \to \mathcal{C}$ be the normalization (resolving the nodal singularity). Then 
$\pi_{\mathcal{C}}^*L = \mathcal{O}_{\mathbb{P}^1}(3)$, so we have a resolution of singularity
\be
\pi_Q: \mathbb{F}_6  = \mathbb{P}\left(\mathcal{O}_{\mathbb{P}^1}(6)\oplus\mathcal{O}_{ \mathbb{P}^1}\right) 
\to Q
\ee
which covers the resolution of singularity $\pi_{\mathcal{C}}:\mathbb{P}^1 \to \mathcal{C}$. 
In Figure \ref{fig:F6}, $\pi_Q^{-1}(q)=\{ q_1, q_2\}$, $f_1$ and $f_2$ are fibers of $\mathbb{F}_6\to \mathbb{P}_1$,
and $b\cdot b =-6$. 

\begin{figure}[h]
\begin{center}
\includegraphics[scale=0.9]{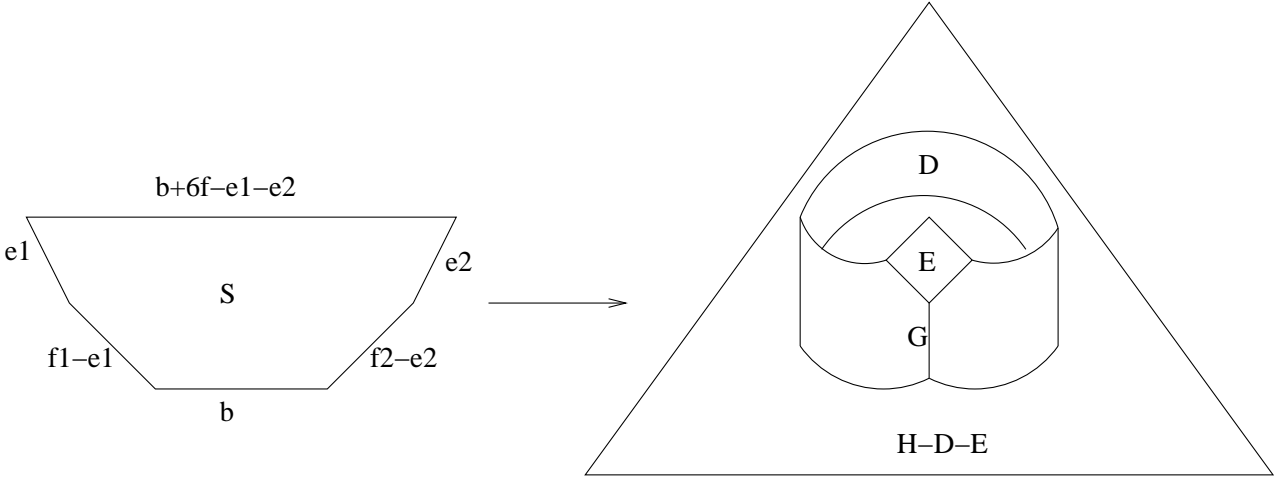}
\caption{Resolution of singularity $\pi_D: S\to D$}
\label{fig:F4}
\end{center}
\end{figure}

Blowing up $\mathbb{F}_6$ at $q_1$ and $q_2$, we obtain a surface $S$ and a resolution
of singularity $\pi_D: S\to D$. In Figure \ref{fig:F4}, $e_1$, $e_2$, $f_1-e_1$, $f_2-e_2$ are -1 curves
in $S$. Under $\pi_D$, $f_1-e_1$ and $f_2-e_2$ are mapped to $G$, and $b$ is mapped to the curve 
$D\cap (H-D-E)$.
Finally, blowing down the -1 curves $f-e_1$ and $f-e_2$ in $S$, we obtain $\mathbb{F}_4\cong D'$. We will demonstrate this explicitly in the next subsection.


\subsection{Topology of $D'$}

To determine the topology of the divisor $D'$ in $\tilde{X}$, note that
\be
c_2 (\tilde{X}) \cdot D' = 8 \,,\qquad 
(D')^3 = D' \cdot ( (H-D' - E') + E' - H)^2 = D' \cdot (10 l - 8 G') = -6 \,.
\ee
Since $K_{\tilde{X}} \cdot D' = 2l - 3 l_0$ this implies
\be
\int c_1 (T_{D'})^2 = 8\,,\qquad \chi_{D'} = 4 \,.
\ee
So in particular, this could be any of the Hirzebruch surfaces $\mathbb{F}_m$, which are fibrations over $\sigma= \mathbb{P}^1$ with fiber class $f$ and intersections
\be
\sigma^2 = -m \,,\qquad \sigma \cdot f = 1\,,\qquad f^2 =0\,.
\ee
The canonical class is 
\be
K_{\mathbb{F}_m} = -2 \sigma - (m+2) f 
\ee
with 
\be
K_{\mathbb{F}_m}^2 = 8 \,,\qquad \chi_{\mathbb{F}_m} = 4 \,,\qquad 
K_{\mathbb{F}_m} \cdot \sigma =  m-2 \,.
\ee
Thus, only from the last relation can be we distinguish the various Hirzebruch surfaces.
By adjunction formula and $\sigma =\mathbb{P}^1$ and thus $K_{\sigma} = \mathcal{O}(-2)$ we obtain
\be
\mathcal{O} (-2) = K_{\sigma} = K_{D'}|_{\sigma} \cdot N_\sigma \,.
\ee
For $\sigma=D'\cap (H-D'-E')$, the normal bundle is determined from
\be
D' \cap (H-D'-E')^2 = -4 \qquad \Rightarrow \qquad N_{\sigma} = \mathcal{O}(-4) \,,
\ee
and thus
\be
K_{D'}|_{\sigma} = \mathcal{O} (2)\,.
\ee
Thereby we obtain that
\be
D' = \mathbb{F}_4 \,.
\ee


\section{Geometry of $X$}

\subsection{Divisors and Intersections in $X$}
\label{app:Ints}

We compute the intersection matrices for our three-fold, $X$, before the flop.  We have three divisor classes, $H$, $D$, and $E$ along with three curve classes $\ell_0$, $\ell$, and $G$.  We know that $H\cdot H = \ell_0$, $H\cdot E=0$, and $H\cdot D = 3(\ell+G)$.  Further, we have that $D\cdot E = 2\ell$.  This immediately gives us the following table
\begin{equation}\begin{array}{c|c|c|c}
 & H & E & D \\ \hline
 H & \ell_0 & 0 & 3(\ell+G) \\
 E & 0 & \ast & 2\ell \\
 D & 3(\ell+G) & 2\ell & \ast
 \end{array}\end{equation}

 We also know some intersections with curves.  For instance, we know that $H\cdot\ell_0=1$, $H\cdot\ell=0$, $H\cdot G=0$.  Because $E$ is an exceptional $\mathbb{P}^1\times\mathbb{P}^1$, we know that the normal bundle of $E$ inside $X$ is $(-1,-1)$ and hence that $\ell\cdot E = -1$.  Further, we know that $E\cdot(\ell+G)=0$ so $G\cdot E=1$.  We also have $E\cdot\ell_0=0$.  Turning now to intersections with $D$, we might think that $\ell\cdot D=0$ because the normal bundle of $\ell=E\cap D$ inside $E$ has degree 0.  However, $\ell$ always intersects the other $\mathbb{P}^1$ inside $E$ at a point and this other $\mathbb{P}^1$ is also identified with a cycle in $D$.   This means that actually $\ell\cdot D=1$.  Next, we know that $G$ is a $(-1,-1)$ curve so that $G\cdot D=-2$.  The factor of 2 arises because $G$ effectively meets $D$ twice.  All of this is consistent with the fact that we expect the restriction of $N_{D/X}$ to $\ell+G$ to have degree $-1$, a condition which is equivalent to $D\cdot (\ell+G)=-1$.  This is all summarized in the following table
 \begin{equation}\begin{array}{c|c|c|c}
 & H & E & D \\ \hline
 \ell_0 & 1 & 0 & 0 \\
 \ell & 0 & -1 & 1 \\
 G & 0 & 1 & -2 \end{array}\end{equation}
 
 To complete our intersection numbers, we study the divisor $H-D-E$.  This is the $\mathbb{P}^2$ that descends from the class $H$ inside $\mathbb{P}^3$ and contains the full nodal curve ${\cal{C}}$.  We now recall that this $\mathbb{P}^2$ fails to intersect $E$
 \begin{equation}(H-D-E)\cdot E = 0\,.
 \end{equation}
Further, $(H-D-E)\cdot D$ is the nodal curve ${\cal{C}}$, which is in the class $3h$ inside $H-D-E$, where $h$ is the hyperplane class of $H-D-E$.  This means that ${\cal{C}}$ has a normal bundle inside $H-D-E$ of degree 9 or, in other words, that
\begin{equation}(H-D-E)\cdot D\cdot D = 9\,.
\end{equation}
This is now enough to fix the remaining intersections, $E^2$ and $D^2$.  For starters, we see that
\begin{equation}E^2 = (H-D)\cdot E = -2\ell\,.
\end{equation}
Further, we can determine $(H-D-E)\cdot D$ from
\begin{equation}\begin{split}\left[(H-D-E)\cdot D\right]\cdot H &= 3\\
\left[(H-D-E)\cdot D\right]\cdot D &= 9\\
\left[(H-D-E)\cdot E\right]\cdot E &= 0\,,
 \end{split}\end{equation}
 where we computed the first and third using known intersection data and the second follows from the reasoning above.  This uniquely fixes
 \begin{equation}
 (H-D-E)\cdot D = 3\left[\ell_0 - 3(\ell+G)\right]
\,. 
\end{equation}
We can understand this result by recalling that $(H-D-E)\cdot D$ is precisely the nodal curve ${\cal{C}}$.  An element of the hyperplane class $h$ inside $H-D-E$ will generically intersect ${\cal{C}}$ three times.  As such, $h\sim \ell_0-3(\ell+G)$.  Now, the above result is simply the statement that ${\cal{C}}$ is in the class $3h$ inside $H-D-E$.

This result finally allows us to compute $D^2$ via
\begin{equation}\begin{split}D^2 &= H\cdot D - E \cdot D - (H-D-E)\cdot D \\
&= 3(\ell+G) - 2\ell - 3\left[\ell_0-3(\ell+G)\right]\\
&= -3\ell_0 +12(\ell+G) - 2\ell \,.
\end{split}\end{equation}
Our complete intersection tables are therefore
\begin{equation}
\begin{array}{c|c|c|c}
& H & E & D \\ \hline
H & \ell_0 & 0 & 3(\ell+G) \\
E & 0 & -2\ell & 2\ell \\
D & 3(\ell+G) & 2\ell & -3\ell_0 + 12(\ell+G) - 2\ell \\
H-D-E & \ell_0-3(\ell+G) & 0 & 3\left[\ell_0-3(\ell+G)\right]
\end{array}
\qquad\qquad
\begin{array}{c|c|c|c}
& H & E & D \\ \hline
\ell_0 & 1 & 0 & 0 \\
\ell & 0 & -1 & 1 \\
G & 0 & 1 & -2
\end{array}\end{equation}

One can check that the triple intersection matrix is associative by directly computing
\begin{equation}\begin{split}
H^3 &= 1 \\
H^2D &= 0 \\
H^2E &= 0 \\
D^3 &= -14 \\
D^2H &= -3 \\
D^2E &= 2 \\
E^3 &= 2 \\
E^2 H &= 0 \\
E^2 D &= -2 \\
HDE &= 0\,.
\end{split}\end{equation}

\subsection{$K_X$ and $c_2(T_X)$}

The canonical class is
 \begin{equation}K_X = -4H + (D+E) + E \,.
 \end{equation}
To determine $c_2(T_X)$, we use the following formula for the Euler character of a divisor $S\subset X$
\be
\chi(S) = S^3 + K_X\cdot S^2 + c_2(T_X)\cdot S \,.
\ee
Given this, we can determine $c_2(T_X)$ provided we already know the topology of three divisors.  Indeed, we know that $E=\mathbb{P}^1\times\mathbb{P}^1$, $H=dP_3$, and $H-D-E=\mathbb{P}^2$.  Recalling that
\begin{equation}
\chi(\mathbb{P}^2)=3\,,\qquad \chi(\mathbb{P}^1\times\mathbb{P}^1)=4\,,
\qquad \chi(dP_n)=3+n\,,
\end{equation}
we can now compute
\begin{equation}\begin{split}
c_2(T_X)\cdot H &= \chi(H) - H^3 - K_X\cdot H^2 = 9\\
c_2(T_X)\cdot E &= \chi(E) - E^3 - K_X\cdot E^2 = 0\\
c_2(T_X)\cdot (H-D-E) &= \chi(H-D-E) - (H-D-E)^3 - K_X\cdot (H-D-E)^2 =-3\,.
\end{split}\end{equation}
This fixes
\begin{equation}
c_2(T_X) = 9\ell_0 - 12(\ell+G)=9H^2 - 4H\cdot D \,.
\end{equation}


\section{Geometry of $\tilde{X}$}

\subsection{Divisors and Intersections in $\tilde{X}$}
\label{app:Xtilde}

After the flop, our three-fold $\tilde{X}$ has divisors $H$, $E'$, and $D'$ along with curve classes $\ell_0$, $\ell$, and $G'$.  We still have that $H\cdot H=\ell_0$ and $H\cdot E'=0$.  For $H\cdot D'$, however, we note that the $\mathbb{P}^1$ corresponding to $\ell+G$ has been replaced after the flop by $\ell-G'$.  This means that $H\cdot D'=3(\ell-G')$.  We also have that $E'\cdot D'=2(\ell-G')$.  This information already allows us to nearly complete the table of divisor intersections 

\begin{equation}\begin{array}{c|c|c|c}
& H & E' & D' \\ \hline
H & \ell_0 & 0 & 3(\ell-G') \\
E' & 0 & \ast & 2(\ell-G') \\
D' & 3(\ell-G') & 2(\ell-G') & \ast
\end{array}\end{equation}

Before determining $D^{\prime\,2}$ and $E^{\prime\,2}$, let us first turn to divisor/curve intersections.  As before, $H.\ell_0=1$ while $H.\ell=0$ and $H.G'=0$.  Further, we know that $\ell-G'$ has degree $-1$ in $E'$ so that $D'\cdot (\ell-G') = -1$.  To compute the rest, we note that $H-D'-E'$ is a $dP_1$ with exceptional curve $G'$.  This follows because $H-D-E$ was a $\mathbb{P}^2$ in $X$ and performing the flop essentially blew up a point in this $\mathbb{P}^2$.  Because $G'$ is a $(-1,-1)$ curve, we see that $(H-D'-E')\cdot G' = -1$ and $E'\cdot G' = -1$.  Since $H\cdot G'=0$ this leads us to conclude that $D'\cdot G'=2$, which also makes sense from figure \ref{fig:GlobalThree}.  Since we already saw that $D'\cdot (\ell-G')=-1$ this means that $D'\cdot\ell=1$.  Finally, we expect that $\ell\cdot E'$ is unaffected by the flop so that $\ell\cdot E' = -1$.  This allows us to fill out the table below.  For comparison, we also display the table for $X$ before the flop.
\begin{equation}\begin{array}{c|c|c|c}
& H & E' & D' \\ \hline
\ell_0 & 1 & 0 & 0 \\
\ell & 0 & -1 & 1 \\
G' & 0& -1 & 2
\end{array}
\qquad\qquad
\begin{array}{c|c|c|c}
& H & E & D \\ \hline
\ell_0 & 1 & 0 & 0 \\
\ell & 0 & -1 & 1 \\
G & 0 & 1 & -2 \end{array}
\end{equation}

Finally, we need to determine $E^{\prime\,2}$ and $D^{\prime\,2}$.  By analogy to our results for $X$, we turn to a study of $H-D'-E'$, which is a $dP_1$ with exceptional curve $G'$.  We note that $H-D'-E'$ intersects $E'$ precisely in $G'$
\begin{equation}(H-D'-E')\cdot E'=G'.\label{hdpepep}\end{equation}
Since $(H-D-E)\cdot D$ was in the class $3h$ inside $H-D-E=\mathbb{P}^2$, we expect that $(H-D'-E')\cdot D'$ is in the class $3h-ne$ 
inside $H'-D'-E'=dP_1$ for some $n$.  In fact, it is easy to see that $n$ must be 2 because  $(H-D'-E')\cdot D'$ intersects the 
exceptional curve $G'$ twice.  This means that the normal bundle of $(H-D'-E')\cdot D'$ inside $D'$ has degree 5 or, in other words,
\begin{equation}(H-D'-E')\cdot D'\cdot D' = 5 \,.
\label{hdpepdpdp}\end{equation}
This is now enough to fix $E^{\prime\,2}$ and $D^{\prime\,2}$.  For $E^{\prime\,2}$, we see from \eqref{hdpepep} that
\begin{equation}E^{\prime\,2} = H\cdot E'-D\cdot E'-G' = 0-2(\ell-G')-G' = G'-2\ell\,.
\end{equation}
For $D^{\prime\,2}$, we can determine it from
\begin{equation}\begin{split}\left[(H-D'-E')\cdot D'\right]\cdot H &= 3\\ 
\left[(H-D'-E')\cdot D'\right]\cdot D' &= 5 \\
\left[(H-D'-E')\cdot D'\right]\cdot E' &= 2\,.
\end{split}\end{equation}
where we computed the first and third using known intersection data and the second is simply \eqref{hdpepdpdp}.  This uniquely fixes
\begin{equation}(H-D'-E')\cdot D' =  3\ell_0 -9\ell + 7G' \,,
\end{equation}
from which we conclude that
\begin{equation}D^{\prime\,2} = -3\ell_0 + 10\ell-8G'\,.
\end{equation}
Our complete intersection tables are therefore
\begin{equation}\begin{array}{c|c|c|c} & H & E' & D' \\ \hline
H & \ell_0 & 0 & 3(\ell-G') \\
E' & 0  & G'-2\ell & 2(\ell-G') \\
D' & 3(\ell-G') & 2(\ell-G') & -3\ell_0+10\ell-8G' \\
H-D'-E' & \ell_0-3(\ell-G') & G' & 3\ell_0-9\ell+7G'
\end{array}
\qquad
\begin{array}{c|c|c|c} & H & E' & D' \\ \hline
\ell_0 & 1 & 0 & 0 \\
\ell & 0 & -1 & 1 \\
G' & 0 & -1 & 2
\end{array}\end{equation}

One can check that the triple intersection matrix is associative by directly computing
\begin{equation}\begin{split}
H^3 &=1\\
H^2 D' &=0\\
H^2 E' &=0\\
D^{\prime\,3} &=-6\\
D^{\prime\,2}H &=-3\\
D^{\prime\,2} E' &=-2\\
E^{\prime\,3} &=1\\
E^{\prime\,2}H &=0\\
E^{\prime\,2} D' &=0\\
H D' E' &=0 \,.\\
\end{split}\end{equation}


\subsection{$K_{\tilde{X}}$ and $c_2(T_{\tilde{X}})$}

The canonical class of $\tilde{X}$ is
\begin{equation}
K_{\tilde{X}}=-4H+(D'+E')+E'\,.
\end{equation}

To determine $c_2(T_{\tilde{X}})$, we will again use the formula for the Euler character of a divisor $S\subset \tilde{X}$
\begin{equation}\chi(S) = S^3 + K_{\tilde{X}}\cdot S^2 + c_2(T_{\tilde{X}})\cdot S \,.
\end{equation}
along with knowledge of the topology of three divisor classes.  In particular, we know that $E'=dP_2$, $H=dP_3$, and $H-D'-E'=dP_1$.  This allows us to compute
\begin{equation}\begin{split}c_2(T_{\tilde{X}})\cdot H &= \chi(H)-H^3-K_{\tilde{X}}\cdot H^2 = 9\\
c_2(T_{\tilde{X}})\cdot E' &= \chi(E') - E^{\prime\,3} - K_{\tilde{X}}\cdot E^{\prime\,2} = 2\\
c_2(T_{\tilde{X}})\cdot (H-D'-E') &= \chi(H-D'-E') - (H-D'-E')^3 - K_{\tilde{X}}\cdot (H-D'-E')^2 = -1\,.
\end{split}\end{equation}
which fixes
\begin{equation}c_2(T_{\tilde{X}}) = 9\ell_0-12\ell+10G'\,.
\end{equation}


\section{Examples of Matter Curves in $E$ and $E'$}
\label{app:curveexamples}

Matter curves in $\tilde{X}$ are determined by the intersection of the zero locus of a particular holomorphic section with $E'=dP_2$.  It is important to correctly identify not only the class of the resulting matter curve in $E'$ but whether or not it is an irreducible curve.  The former follows directly from the intersection data of section \ref{subsec:CompactBF} but the latter requires a study of the sections themselves.  In this Appendix, we describe several examples, some of which are useful in the explicit constructions of this paper.

The easiest way to study matter curves in $E'$ is to first understand them in $E$, before the transition from $X$ to $\tilde{X}$, and then carry them through the flop.  Recall that $E=\mathbb{P}^1\times\mathbb{P}^1$ is realized as the submanifold of $\mathbb{P}^3_W$ satisfying
\begin{equation}W_1W_2=W_3W_4\,,
\label{P1P1}\end{equation}
where the $W_i$ are homogeneous coordinates on $\mathbb{P}^3_W$.  We recall that each $W_i$ can be extended globally inside $X$ to a well-defined section of a holomorphic line bundle.  For $W_1$, $W_2$, and $W_3$ the relevant line bundle is ${\cal{O}}(H-E)$ while for $W_4$ it is ${\cal{O}}(3H-D-2E)$.  Because $W_4$ is distinguished from the others, it is necessary to introduce extra sections in order to extend homogeneous equations such as \eqref{P1P1} globally on $X$.  This is easily accomplished by recalling that both $V_0$, which is a section of ${\cal{O}}(H-D-E)$, and $Z_4$, which is a section of ${\cal{O}}(H)$, take the value 1 in the neighborhood of $E$.  As such, any homogeneous equation in the $W_i$ can be extended globally provided we replace $W_i$ with $Z_4V_0W_i$ for $i\ne 4$.  In the rest of this Appendix, we work only locally near $E$ or $E'$ so that homogeneous polynomials in the $W_j$ will suffice.

The middle homology of $E$ is quite simple as it is generated by two curve classes, $\ell_1$ and $\ell_2${\footnote{By construction, $\ell_1$ and $\ell_2$ are homologous inside $X$, but  we are interested in studying them as elements of $H_2(E,\mathbb{Z})$ in this Appendix so we will continue to distinguish them.}},  corresponding to the two different $\mathbb{P}^1$'s.  The intersection product is
\begin{equation}\ell_1^2=\ell_2^2=0\,, \qquad \ell_1\cdot \ell_2=1\,.
\end{equation}

To study $E'$, we note that, from the perspective of $E$, the flop transition corresponds to blowing up a single point $p_0$,
\begin{equation}p_0=[W_1,W_2,W_3,W_4]=[0,0,0,1]\,,
\label{p0def}\end{equation}
inside $E$.  This adds a new exceptional curve, $G'$, with intersections
\begin{equation}
G'\cdot \ell_i=0\,,\qquad G^{\prime\,2}=-1\,.
\end{equation}
The relation to the usual basis of $H_2(dP_2,\mathbb{Z})$ is now as in section \ref{subsubsec:GeomSGUT}
\begin{equation}\begin{split}\ell_1 & \ \leftrightarrow\ \tilde{h}-\tilde{e}_1 \\
\ell_2 & \ \leftrightarrow\  \tilde{h}-\tilde{e}_2 \\
G' &\ \leftrightarrow \ \tilde{h}-\tilde{e}_1-\tilde{e}_2\,.
\end{split}\end{equation}
This means, for instance, that a representative of the class $\ell_1\in H_2(E,\mathbb{Z})$, which does not contain $p_0$ will carry over to $\tilde{h}-\tilde{e}_1$ after the flop.  If $\ell_1$ passes through $p_0$, though, it will carry over to $\ell_1-G'=\tilde{e}_2$ inside $E'$.  To avoid confusion, we will always use the standard basis $\tilde{h}$, $\tilde{e}_1$, and $\tilde{e}_2$ when writing elements of $H_2(E',\mathbb{Z})$ so that $\ell_1,\ell_2$ always refer to elements of $H_2(E,\mathbb{Z})$.

\subsection{Linear Polynomials}

We now study a number of examples involving linear polynomials in the $W_i$.

\subsubsection{Example 1: $W_2=0$}  As a first example, let us consider the curve $W_2=0$.
In $E$, this is the intersection of $W_2=0$ and \eqref{P1P1} inside $\mathbb{P}^3_W$, which has two components
\begin{equation}W_2=W_3=0\,,\qquad W_2=W_4=0\,.
\end{equation}
We now fix our identifications of $\ell_1$ and $\ell_2$ so that
\begin{equation}\begin{split}
W_2=W_3=0 &\quad\leftrightarrow_E \quad  \ell_2 \\
W_2=W_4=0 &\quad  \leftrightarrow_E \quad \ell_1\,.
\end{split}\end{equation}
To follow this through the flop, we note that $W_2=W_3=0$ contains $p_0$ while $W_2=W_4=0$ does not.  This means that
\begin{equation}\begin{split}W_2=W_3=0 &\quad \leftrightarrow_{E'} \quad \tilde{e}_1 \\
W_2=W_4=0 & \quad \leftrightarrow_{E'} \quad \tilde{h}-\tilde{e}_1 \,.
\end{split}\end{equation}
We therefore see that $W_2=0$ splits into two irreducible components.  The net class is $(\tilde{e}_1)+(\tilde{h}-\tilde{e}_1)=\tilde{h}$, as expected from section \ref{subsec:CompactBF} and the fact that $W_2$ extends to a section of ${\cal{O}}(H-E')$.


\subsubsection{Example 2: $W_4=0$}
For a second example, consider the curve $W_4=0$.  Inside $E$, this also has two components
\begin{equation}\begin{split}W_1=W_4=0 &\quad \leftrightarrow_E \quad  \ell_2 \\
W_2=W_4=0 &\quad \leftrightarrow_E\quad  \ell_1 \,.
\end{split}\end{equation}
Each of these misses $p_0$ so they carry over after the flop to 
\begin{equation}\begin{split}
W_1=W_4=0 &\quad \leftrightarrow_{E'} \quad \tilde{h}-\tilde{e}_2 \\
W_2=W_4=0 &\quad \leftrightarrow_{E'} \quad \tilde{h}-\tilde{e}_1\,.
\end{split}\end{equation}
The total class of $W_4=0$ is $2\tilde{h}-\tilde{e}_1-\tilde{e}_2$, as expected from section \ref{subsec:CompactBF} and the fact that $W_4$ extends to a section of ${\cal{O}}(3H-D'-2E')$.


\subsubsection{Example 3: $W_3=W_4$} 
To see that one can also obtain irreducible matter curves from linear polynomials{\footnote{Indeed, generic linear polynomials give rise to irreducible matter curves.}}, we consider now the example $W_3=W_4$.  Before the flop, this is an irreducible curve in the class $\ell_1+\ell_2\in H_2(E,\mathbb{Z})$.  Because it misses $p_0$, it carries over to a curve in the class $2\tilde{h}-\tilde{e}_1-\tilde{e}_2\in H_2(E',\mathbb{Z})$.  This is consistent with what we expect from section \ref{subsec:CompactBF} and the fact that an extension of $W_3=W_4$ to $\tilde{X}$ takes the form $Z_4V_0W_3=W_4$, which is a section of ${\cal{O}}(3H-D'-2E')$.

We now list several common curves and their classes in $E'$
\begin{equation}\begin{array}{c|c}\text{Equation} & \text{Class in $dP_2$} \\ \hline
W_2=W_3=0 & \tilde{e}_1 \\
W_1=W_3=0 & \tilde{e}_2 \\
W_2=W_4=0 & \tilde{h}-\tilde{e}_1 \\
W_1=W_4=0 & \tilde{h}-\tilde{e}_2 \\
W_3=W_4 & 2\tilde{h}-\tilde{e}_1-\tilde{e}_2 \\
W_1=W_2 & \tilde{h}
\end{array}\end{equation}


\subsection{A Degenerate $T^2$ Example}
\label{app:degT2}

We now turn to more intricate examples involving higher degree polynomials that are useful for the explicit constructions in this paper.  

One such example is the quadratic polynomial
\begin{equation}
W_1(W_4-W_3-W_2)+W_2W_4\,.
\label{quadpoly}\end{equation}
The extension of this polynomial to $\tilde{X}$ takes the form
\begin{equation}W_1(W_4-V_0Z_4W_3-V_0Z_4W_2)+W_2W_4\,,
\end{equation}
and is a section of ${\cal{O}}(4H-D'-3E')$.  From section \ref{subsec:CompactBF}, this means that its intersection with $E'$ will be an element of the class $3\tilde{h}-\tilde{e}_1-\tilde{e}_2$.  Irreducible curves in the class $3\tilde{h}-\tilde{e}_1-\tilde{e}_2$ have genus 1 so generic polynomials of this type describe $T^2$'s.  By inspection, however, we see that \eqref{quadpoly} contains an irreducible component
\begin{equation}W_1=W_4=0\,,
\label{he2comp}\end{equation}
in the class $\tilde{h}-\tilde{e}_2$.  This means that this particular curve is in fact a torus that has degenerated into multiple components.  The remaining component is a curve in the class $2\tilde{h}-\tilde{e}_1$, a parametrization of which in terms of an affine coordinate $x$ is given by
\begin{equation}[W_1,W_2,W_3,W_4]=[x(x^2+x-1),x,x^2,x^2+x-1]\,.
\label{t2comp}\end{equation}
From this it is easy to see that, as expected, the $2\tilde{h}-\tilde{e}_1$ component, \eqref{t2comp}, intersects the $\tilde{h}-\tilde{e}_2$ component \eqref{he2comp}, at precisely two points, namely the two roots of $x^2+x-1$, $x_{\pm}=-\frac{1}{2}\pm \frac{\sqrt{5}}{2}$.  The "point at infinity" of the $2\tilde{h}-\tilde{e}_1$ component that is not included in the parametrization \eqref{t2comp} is simply $[1,0,0,0]$.

\section{Several Classes of Elliptic Fibration}
\label{app:Fibs}

In this Appendix, we describe several classes of elliptic fibrations in which the holomorphic section $P$ that determines the $\mathbf{5}$ matter curves factorizes into a product of several sections
\begin{equation}P=\prod_i P_i \,,
\label{genPans}\end{equation}
each of which we will eventually choose to meet $S_{\rm GUT}$ in a class of relatively low genus.  For generic $P_i$, the divisor defined by $P_i=0$ will meet $S_{\rm GUT}$ in an irreducible curve class.  However, it is possible that such a matter curve is further reducible inside $S_{\rm GUT}$.  For clarity, we will use the term "factors" in what follows for the $P_i$, which may or may not be irreducible inside $S_{\rm GUT}$.  We will reserve the term ``component" for an irreducible curve class.

Before turning to a detailed discussion of $P$, let us first focus on $h$, which is a section of ${\cal{O}}(4H-D'-2E')$.  In order to account for the $-D'$, we must include a factor of $W_4$, $V_0$, or $V_1$.  As such, the most general expression for $h$ that we can write is
\begin{equation}h^{(4H-D'-2E')} = h_0^{(3H-E')}V_0^{(H-D'-E')} + h_1^{(H-E')} V_1^{(3H-D'-E')} + h_2^{(H)}W_4^{(3H-D'-2E')}\,,
\end{equation}
where $h_0$, $h_1$, and $h_2$ are sections of the indicated bundles.  In general, it is simplest to choose any section of a bundle of the form $(n+m)H-mE'$ to be proportional to $Z_4^n$.  In that case, what remains is a section of $m(H-E')$, which we can choose to be any degree $m$ polynomial $F_m(W_1,W_2,W_3)$.  Because $Z_4=1$ on $S_{\rm GUT}$, the restriction of such a section to $S_{\rm GUT}$ is easily determined as in Appendix \ref{app:curveexamples} by following the intersection of $F_m(W_1,W_2,W_3)$ with $W_1W_2=W_3W_4$ inside $\mathbb{P}^3_W$ through the flop.

A particularly simple choice for $h$ that we will use in the following is
\begin{equation}h^{(4H-D'-2E')}=Z_4^{(H)}\left(W_4^{(3H-D'-2E')}-Z_4^{(H)}V_0^{(H-D'-E')}A^{(H-E')}\right)\,.
\end{equation}
For a generic linear polynomial $A$ in $W_1$, $W_2$, and $W_3$, the restriction of $h$ to $S_{\rm GUT}$ will be an irreducible curve in the class $2\tilde{h}-\tilde{e}_1-\tilde{e}_2$.  Because $h$ determines the $\mathbf{10}$ matter curve, $\Sigma_{10}$, this means that in all examples in this Appendix we have
\begin{equation}
\Sigma_{10} = 2\tilde{h}-\tilde{e}_1-\tilde{e}_2\,.
\end{equation}

Let us now return to the issue of $P$.  When we build our fibration we do not directly specify $P$ but rather the sections $h$, $q$, $H$, $f_3$, and $g_5$ which determine $P$ via \eqref{PQdef}
\begin{equation}\begin{split}
P^{(32H-8D'-21E')}&=-3H^{(8H-2D'-5E')}\left(q^{(12H-3D'-8E')}\right)^2-3h^{4H-D'-2E')}f_3^{(16H-4D'-11E')}q^{(12H-3E'-8E')}\\
&\qquad +2g_5^{(24H-6D'-17E')}\left(h^{(4H-D'-2E')}\right)^2\,.
\end{split}\label{Peqnapp}\end{equation}
It is therefore crucial to demonstrate that a particular choice of these sections can be made to obtain $P$ of a particular desired form \eqref{genPans}.  Our strategy for doing this will be to study these sections in a series expansion in $h$.  In particular, if we can demonstrate that the expansion of $P+3Hq^2+3hf_3q$ begins at order $h^2$ for a particular choice of $P$, then we can always choose $g_5=(2h)^{-2}\left(P+3Hq^2+3hf_3q\right)$.

To proceed, then, we need to write the sections $H$, $q$, and $f_3$ as
\begin{equation}\begin{split}
H^{(8H-2D'-5E')} &= H_0^{(8H-2D'-5E')}+h^{(4H-D'-2E')}H_1^{(4H-D'-3E')}+\cdots \\
q^{(12H-3D'-8E')} &= q_0^{(12H-3D'-8E')} + h^{(4H-D'-2E')}q_1^{(8H-2D'-6E')} + \cdots \\
f_3^{(16H-4D'-11E')} &= f_{3,0}^{(16H-4D'-11E')}+\cdots \,.\\
\end{split}\end{equation}
We also need to expand our ansatz for $P$ according to
\begin{equation}P^{(32H-8D'-21E')} = P_0^{(32H-8D'-21E')} + h^{(4H-D'-2E')}P_1^{(28H-7D'-19E')} + \cdots\,.
\label{Pfactexp}\end{equation}
where we explicitly write the bundles associated to various coefficients.  Expanding $P+3Hq^2+3hf_3q$, we find
\begin{equation}P+3Hq^2+3hf_3q = \left(P_0+3H_0q_0^2\right)+ h\left[P_1+3q_0\left(2q_1H_0+q_0H_1+f_{3,0}\right)\right]+O(h^2)\,.
\label{PHqexp}\end{equation}
For this to vanish at order $h^0$ we need to choose $H_0$ and $q_0$ so that
\begin{equation}\boxed{P_0+3H_0q_0^2=0}\,.
\label{Pcond1}\end{equation}
In other words, we must choose the factors of $H_0$ and $q_0$ to correspond to various factors of $P_0$.  Because $SO(12)$ points occur when $P_0=q_0=0$ and $E_6$ points occur when $P_0=H_0=0$, this essentially determines the structure of all matter curve intersections that we need for the MSSM.  

As for the $h^1$ term in \eqref{PHqexp}, we can make a choice of $f_{3,0}$ that eliminates it provided
\begin{equation}\boxed{P_1 = q_0\tilde{P}_1}\,.
\label{Pcond2}\end{equation}

Given an ansatz for $P$, then, we can realize it in an explicit elliptic fibration provided a choice for $H_0$ and $q_0$ exists that satisfies \eqref{Pcond1} and \eqref{Pcond2}.  We now turn to the study of several examples.


\subsection{Example 1: $P$ factors into curves of $g=0$}
\label{app:Ex1}

We first consider a ``maximal" factorization of $P$ into a product of factors $P_i$ that each meet $S_{\rm GUT}$ in a curve class of genus 0.  Because $P$ is a section of $32H-8D'-21E'$, we must include factors of $W_4$, $V_0$, and $V_1$ to account for the $-8D'$ part.  Since we used only $W_4$ and $V_0$ in writing $h$ above, it is simplest to use only these two sections for $P$ as well.  We therefore seek to obtain
\begin{equation}P^{(32H-8D'-21E')}_{\text{Ex 1}} = \left(Z_4^{(H)}\right)^3F_p^{(5H-5E')}\prod_{i=1}^8\left(W_4^{(3H-D'-2E')}-Z_4^{(H)}V_0^{(H-D'-E')}G_i^{(H-E')}\right)\,.
\label{Pex1}\end{equation}
This will correspond to splitting the $\mathbf{5}$ matter curve, which is specified by $P$, according to
\begin{equation}\Sigma_5=21\tilde{h}-8(\tilde{e}_1+\tilde{e}_2)\rightarrow 8\times (2\tilde{h}-\tilde{e}_1-\tilde{e}_2) + 5\times \tilde{h}\,.
\end{equation}

To check the conditions \eqref{Pcond1} and \eqref{Pcond2}, it is necessary to expand our particular ansatz \eqref{Pex1} as in \eqref{Pfactexp}.  In the notation of \eqref{Pfactexp} we find
\begin{equation}\begin{split}P_0^{(32H-8D'-21E')} &= \left(Z_4^{(H)}\right)^{11}\left(V_0^{(H-D'-E')}\right)^8F_p^{(5H-5E')}s_8\left(G_i^{(H-E')}-A^{(H-E')}\right) \\
P_1^{(28H-7D'-19E')} &= \left(Z_4^{(H)}\right)^9\left(V_0^{(H-D'-E')}\right)^7F_p^{(5H-5E')}s_7\left(G_i^{(H-E')}-A^{(H-E')}\right)\,.
\end{split}\end{equation}
where $s_n(x_i)$ denote Schur polynomials, which can be defined by
\begin{equation}\prod_{i=1}^N\left(x+x_i\right) = \sum_{n=0}^N x^{N-n}s_n(x_i)\,.
\end{equation}
We can now satisfy both \eqref{Pcond1} and \eqref{Pcond2} while avoiding higher order zeros in $H_0$ and $q_0$ by choosing
\begin{equation}\begin{split}F_p^{(5H-5E')} &= \prod_{i=1}^5\left(G_i^{(H-E')}-A^{(H-E')}\right) \\
H_0^{(8H-2D'-5E')} &= -\frac{1}{3}\left(Z_4^{(H)}\right)^3\left(V_0^{(H-D'-E')}\right)^2\prod_{j=6}^8\left(G_j^{(H-E')}-A^{(H-E')}\right)\\
q_0^{(12H-3D'-8E')} &= \left(Z_4^{(H)}\right)^4\left(V_0^{(H-D'-E')}\right)^3 F_p^{(5H-5E')}\,.
 \\
\end{split}\end{equation}

As alluded to above, these choices completely fix the structure of matter curves and their intersections.  To write the various factors of the $\Sigma_5$ matter curve, it is helpful to split the $G_i$ into two groups.  In particular, we define $G_a=G_i$ for $a=1,\ldots 5$ and $\tilde{G}_{\mu}=G_{i+5}$ for $\mu=6,\ldots,8$.  With this notation, we can write the $\Sigma_5$ matter curves as
\begin{equation}\begin{array}{c|c|c}\text{Matter Curve} & \text{Equation} & \text{Class in }dP_2 \\ \hline
\Sigma_{5,a}\,\, (a=1,\ldots,5) & W_4 - Z_4V_0 G_a & 2\tilde{h}-\tilde{e}_1-\tilde{e}_2 \\
\Sigma_{5,A}\,\, (A=1,\ldots,3) & W_4 - Z_4V_0 \tilde{G}_A & 2\tilde{h}-\tilde{e}_1-\tilde{e}_2 \\
\Sigma_{5,\mu}\,\, (\mu=1,\ldots,5) & G_{\mu}-A & \tilde{h}
\end{array}\label{mattertable}\end{equation}
Each of these matter curves corresponds to the intersection of $S_{\rm GUT}$ with a divisor in $\tilde{X}$.  It is possible that this intersection is further reducible inside $S_{\rm GUT}$ and, indeed, as described in the text this will be necessary to engineer Higgs fields.

The intersections of the $\mathbf{5}$ matter curves of \eqref{mattertable} with the $\mathbf{10}$ matter curve have the following generic structure
\begin{equation}\begin{array}{c|c} \text{Intersecting Curves} & \text{Singularity Type} \\ \hline
\Sigma_{10}\cap \Sigma_{5,a} \cap \Sigma_{5,\mu}\,\,\,(a=\mu) & SO(12) \\
\Sigma_{10}\cap \Sigma_{5,A} & E_6
\end{array}\end{equation}
Unfortunately, no single factor of $\Sigma_5$ participates in both an $SO(12)$ and an $E_6$ point.  This is problematic because the only way to get both $H_u$ and $H_d$ without extra exotics is to realize them on two components of a single reducible factor of $P$.  The absence of a single factor that participates in both types of Yukawa couplings means that we either have to abandon this family of constructions or realize $H_u$ and $H_d$ on different factors, in which case we will have to live with extra exotics.






\subsection{Example 2: $P$ factors into many curves of $g=0$ and one with $g=1$}
\label{app:Ex2}

In the previous example, we ran into trouble because $P$ was ``too factorized" in the sense that it was broken into so many factors that no single one contained both an $SO(12)$ and an $E_6$ enhancement point.  To rectify the situation, let us consider a slightly less factorized form in which one of the factors corresponds to a curve class of genus 1.  If we succeed in writing such a family of solutions then we might hope to realize $H_u$ and $H_d$ on a representative of this class in which the $T^2$ has degenerated to two individual $\mathbb{P}^1$'s.

More specifically, we modify our previous ansatz for $P$ to
\begin{equation}\begin{split}
P_{\text{Ex 2}}^{(32H-8D'-21E')} &= \left(Z_4^{(H)}\right)^2\tilde{F}_p^{(4H-4E')}\prod_{i=1}^7\left(W_4^{(3H-D'-2E')}-Z_4^{(H)}V_0^{(H-D'-E')}G_i^{(H-E')}\right)\times \\
\times& \left[Z_4^{(H)}B^{(H-E')}\left(W_4^{(3H-D'-2E')}-Z_4^{(H)}V_0^{(H-D'-E')}{\cal{G}}^{(H-E')}\right)+h^{(4H-D'-2E')}C^{(H-E')}\right] \,.
\end{split}\label{Pex2}\end{equation}
This corresponds to splitting $\Sigma_5$ according to
\begin{equation}\Sigma_5=21\tilde{h}-8(\tilde{e}_1+\tilde{e}_2) \rightarrow (3\tilde{h}-\tilde{e}_1-\tilde{e}_2)+7\times\left(2\tilde{h}-\tilde{e}_1-\tilde{e}_2\right) + 4\times \tilde{h}\,.
\end{equation}
To check \eqref{Pcond1} and \eqref{Pcond2} we now expand \eqref{Pex2} as in \eqref{Pfactexp} with
\begin{equation}\begin{split}
P_0^{(32H-8D'-21E')} &= \left(Z_4^{(H)}\right)^{11}\left(V_0^{(H-D'-E')}\right)^{8}\tilde{F}_p^{(4H-4E')}B^{(H-E')}\times\\
& \quad \times s_8\left(A^{(H-E')}-G_i^{(H-E')},A^{(H-E')}-{\cal{G}}^{(H-E')}\right) \\
P_1^{(28H-7D'-19E')} &= \left(Z_4^{(H)}\right)^9\left(V_0^{(H-D'-E')}\right)^7 \tilde{F}_p^{(4H-4E')}\times\\
&\qquad\times \left[B^{(H-E')}s_7\left(A^{(H-E')}-G_i^{(H-E')},A^{(H-E')}-{\cal{G}}^{(H-E')}\right)\right.\\
&\qquad\qquad\left.+C^{(H-E')}s_7\left(A^{(H-E')}-G_i^{(H-E')}\right)\right]\,.
\end{split}\end{equation}
It is now easy to see that both \eqref{Pcond1} and \eqref{Pcond2} can be satisfied if we take
\begin{equation}\begin{split}
\tilde{F}_p^{(4H-4E')}&=\prod_{i=1}^4\left(G_i^{(H-E')}-A^{(H-E')}\right) \\
B^{(H-E')} &= G_5^{(H-E')}-A^{(H-E')} \\
H_0^{(8H-2D'-5E')} &= -\frac{1}{3}\left(Z_4^{(H)}\right)^3\left(V_0^{(H-D'-E')}\right)^2\left({\cal{G}}^{(H-E')}-A^{(H-E')}\right)\prod_{j=6}^7\left(G_j^{(H-E')}-A^{(H-E')}\right) \\
q_0^{(12H-3D'-8E')} &= \left(Z_4^{(H)}\right)^4\left(V_0^{(H-D'-E')}\right)^3\tilde{F}_p^{(4H-4E')}B^{(H-E')}\,.
\end{split}\end{equation}

To describe the various matter curves and their intersections, it is again useful to break apart the $G_i$ into groups.  This time, we write $G_a=G_i$ for $i=1,\ldots,4$ and $\tilde{G}_{A}=G_{A+5}$ for $A=1,2$.  With this notation, we can write the $\Sigma_5$ matter curves as
\begin{equation}\begin{array}{c|c|c}\text{Matter Curve} & \text{Equation} & \text{Class in $dP_2$} \\ \hline
\Sigma_{5,a}\,\,\, (a=1,\ldots,4) & W_4-Z_4V_0G_a & 2\tilde{h}-\tilde{e}_1-\tilde{e}_2 \\
\tilde{\Sigma}_{5,\mathbb{P}^1} & W_4 - Z_4V_0 G_5 & 2\tilde{h}-\tilde{e}_1-\tilde{e}_2 \\
\Sigma_{5,A}\,\,\, (A=1,2) & W_4 - Z_4V_0\tilde{G}_{A} & 2\tilde{h}-\tilde{e}_1-\tilde{e}_2 \\
\Sigma_{5,\mu}\,\,\, (\mu=1,\ldots,4) & G_{\mu}^{(H-E')}-A^{(H-E')} & \tilde{h} \\
\tilde{\Sigma}_{5,T^2} & Z_4B(W_4-Z_4V_0{\cal{G}})+hC & 3\tilde{h}-\tilde{e}_1-\tilde{e}_2
\end{array}\label{ex2mattercurves}\end{equation}

The intersections of the $\mathbf{5}$ matter curves of \eqref{ex2mattercurves} with the $\mathbf{10}$ matter curve have the following generic structure
\begin{equation}\begin{array}{c|c}\text{Intersecting Curves} & \text{Singularity Type} \\ \hline
\Sigma_{10}\cap \Sigma_{5,a}\cap \Sigma_{5,\mu}\,\,\,(\mu=a) & SO(12) \\
\Sigma_{10}\cap \tilde{\Sigma}_{5,\mathbb{P}^1}\cap \tilde{\Sigma}_{5,T^2} & SO(12) \\
\Sigma_{10} \cap \Sigma_{5,A} & E_6 \\
\Sigma_{10} \cap \tilde{\Sigma}_{5,T^2} & E_6
\end{array}\end{equation}

\subsubsection{Candidate Curves for MSSM Matter}

Quite nicely, $\tilde{\Sigma}_{5,T^2}$ participates in both the $SO(12)$ and $E_6$ Yukawa couplings.  Indeed, we can get all of the desired Yukawa couplings if we engineer all three generations of $\mathbf{\overline{5}}_M$ on $\tilde{\Sigma}_{5,\mathbb{P}^1}$ and both $H_u$ and $H_d$ on $\tilde{\Sigma}_{5,T^2}$.  To do the latter, we will need a nongeneric choice of $C$, $G_5$, and ${\cal{G}}$ so that $\tilde{\Sigma}_{5,T^2}$ intersects $S_{\rm GUT}$ in a reducible curve that is the sum of two $\mathbb{P}^1$'s, each of which exhibits a nontrivial restriction of the hypercharge flux $[F_Y]\sim \tilde{e}_1-\tilde{e}_2$.

To close this subsection, let us demonstrate that with a particular choice of $A^{(H-E')}$, $B^{(H-E')}$, $C^{(H-E')}$, ${\cal{G}}^{(H-E)}$, and $G_5^{(H-E')}$ it is possible to arrange for all of the Yukawa structures described in section \ref{sec:Loc}.  In particular, we want $\tilde{\Sigma}_{5,T^2}$ to split in $S_{\rm GUT}$ into two components such that one meets $\tilde{\Sigma}_{5,\mathbb{P}^1}$ and $\Sigma_{10}$ in exactly one $SO(12)$ point and the other meets $\Sigma_{10}$ in exactly one $E_6$ point.  We would then need to engineer $H_d$ on the former and $H_u$ on the latter.

For $\tilde{\Sigma}_{5,T^2}$, let us choose 
\begin{equation}\begin{split}
A^{(H-E')} &= W_3^{(H_E')} \\
B^{(H-E')} &= W_2^{(H-E')} \\
{\cal{G}}^{(H-E')} &= W_1^{(H-E')} \\
C^{(H-E')} &= W_1^{(H-E')} \\
G_5^{(H-E')} &= W_3^{(H-E')}+W_2^{(H-E')}\,.
\end{split}\end{equation}
With these choices, the matter curves $\Sigma_{10}$, $\tilde{\Sigma}_{5,T^2}$, and $\tilde{\Sigma}_{5,\mathbb{P}^1}$ are given by the equations
\begin{equation}\begin{array}{c|c}\text{Curve} 
& \text{Equation}\\ \hline
\Sigma_{10} & W_4-W_3 \\
\tilde{\Sigma}_{5,T^2} & W_2(W_4-W_1)+W_1(W_4-W_3) \\
\Sigma_{5,\mathbb{P}^1} & W_4-W_3-W_2
\end{array}\end{equation}
Note that $\tilde{\Sigma}_{5,T^2}$ is precisely the curve that we study in section \ref{app:degT2}.  In particular, it is a reducible curve in the class $3\tilde{h}-\tilde{e}_1-\tilde{e}_2$ which splits into components $(\tilde{h}-\tilde{e}_2)$ and $2\tilde{h}-\tilde{e}_1$.  Using the description of this curve in section \ref{app:degT2}, it is easy to work out its intersections with $\Sigma_{10}$ and $\tilde{\Sigma}_{5,\mathbb{P}^1}$
\begin{equation}\begin{array}{c|c|c}
\text{Intersecting Curves} & \text{Intersection Point} & \text{Singularity Type} \\ \hline
\Sigma_{10}\cap \tilde{\Sigma}_{5,T^2}^{(2\tilde{h}-\tilde{e}_1)}\cap\tilde{\Sigma}_{5,\mathbb{P}^1} 
										& [1,0,0,0] & SO(12) \\
\Sigma_{10}\cap \tilde{\Sigma}_{5,T^2}^{(2\tilde{h}-\tilde{e}_1)} & [1,1,1,1] & E_6\\
\Sigma_{10}\cap \tilde{\Sigma}_{5,T^2}^{(\tilde{h}-\tilde{e}_2)} & [0,1,0,0] & E_6
\end{array}\end{equation}


\subsection{Example 3: $P$ factors into many curves of genus 0 and one of genus 1}
\label{app:Ex3}

It is easy to write a different ansatz in which $P$ splits into a sum of genus 0 factors and a genus 1 factor.
\begin{equation}\begin{split}
&P_{\text{Ex 3}}^{(32H-8D'-21E')} =\\
& \left(Z_4^{(H)}\right)^3F_p^{(5H-5E')}\prod_{i=1}^6\left(W_4^{(3H-D'-2E')}-Z_4^{(H)}V_0^{(H-D'-E')}G_i^{(H-E')}\right)\times \\
&\times \left[\left(W_4^{(3H-D'-2E')}-Z_4^{(H)}V_0^{(H-D'-E')}G_7^{(H-E')}\right)\left(W_4^{(3H-D'-2E')}-Z_4^{(H)}V_0^{(H-D'-E')}G_8^{(H-E')}\right)\right.\\
&\qquad\qquad\qquad\left.+ h^{(4H-D'-2E')}V_0^{(H-D'-E')}K^{(H-E')}\right]\,.
\end{split}\label{Pex3}\end{equation}
This corresponds to splitting $P$ according to
\begin{equation}P=(4\tilde{h}-2\tilde{e}_1-2\tilde{e}_2)+6\times (2\tilde{h}-\tilde{e}_1-\tilde{e}_2) + 5\times \tilde{h}\,.
\end{equation}
To check \eqref{Pcond1} and \eqref{Pcond2}, we expand \eqref{Pex3} as in \eqref{Pfactexp} with
\begin{equation}\begin{split}P_0^{(32H-8D'-21E')} &= \left(Z_4^{(H)}\right)^{11}\left(V_0^{(H-D'-E')}\right)^8F_p^{(5H-5E')}s_8\left(A^{(H-E')}-G_i^{(H-E')}\right) \\
P_1^{(28H-7D'-19E')} &= \left(Z_4^{(H)}\right)^9\left(V_0^{(H-D'-E')}\right)^7\\
&\qquad\times \left[F_p^{(5H-5E')}s_7\left(A^{(H-E')}-G_i^{(H-E')}\right)+ K^{(H-E')}\prod_{i=1}^6\left(A^{(H-E')}-G_i^{(H-E)}\right)\right]
\end{split}\end{equation}
It is easy to see that both \eqref{Pcond1} and \eqref{Pcond2} can be satisfied if we take
\begin{equation}\begin{split}
F_p^{(5H-5E')} &= \left(A^{(H-E')}-G_7^{(H-E)}\right)\prod_{i=1}^4\left(A^{(H-E')}-G_i^{(H-E)}\right) \\
K^{(H-E')} &= \left(A^{(H-E')}-G_7^{(H-E')}\right) \\
H_0^{(8H-2D'-5E')} &= -\frac{1}{3}\left(Z_4^{(H)}\right)^3\left(V_0^{(H-D'-E')}\right)^2\left(A^{(H-E')}-G_8^{(H-E')}\right)\prod_{j=5}^6\left(A^{(H-E')}-G_i^{(H-E')}\right) \\
q_0^{(12H-3D'-8E')} &= \left(Z_4^{(H)}\right)^4\left(V_0^{(H-D'-E')}\right)^3F_p^{(5H-5E')}\,.
\end{split}\end{equation}

We can list the $\mathbf{5}$ matter curves as
\begin{equation}\begin{array}{c|c|c}\text{Matter Curve} & \text{Equation} & \text{Class in }dP_2 \\ \hline
\Sigma_{5,i}\,\,\,(i=1,\ldots,6) & W_4-Z_4V_0G_i & 2\tilde{h}-\tilde{e}_1-\tilde{e}_2 \\
\Sigma_{5,a}\,\,\,(a=1,\ldots,4) & A-G_a & \tilde{h} \\
\tilde{\Sigma}_{5,\mathbb{P}^1} & A-G_7 & \tilde{h} \\
\tilde{\Sigma}_{5,T^2} & (W_4-Z_4V_0G_7)(W_4-Z_4V_0G_8)+hV_0(A-G_7) & 4\tilde{h}-2(\tilde{e}_1+\tilde{e}_2)\,.
\end{array}\label{5mattex3}\end{equation}

The intersections of the $\mathbf{5}$ matter curves of \eqref{5mattex3} with the $\mathbf{10}$ matter curve have the following generic structure
\begin{equation}\begin{array}{c|c}\text{Intersecting Curves} & \text{Singularity Type} \\ \hline
\Sigma_{10}\cap \Sigma_{5,i}\cap \Sigma_{5,a}\,\,\,(i=a) & SO(12) \\
\Sigma_{10}\cap \tilde{\Sigma}_{5,\mathbb{P}^1}\cap \tilde{\Sigma}_{5,T^2} & SO(12) \\
\Sigma_{10} \cap \Sigma_{5,i}\,\,\,(i=5,6) & E_6 \\
\Sigma_{10}\cap \tilde{\Sigma}_{5,T^2} & E_6 \\
\end{array}\end{equation}


\newpage

\bibliographystyle{JHEP}

\providecommand{\href}[2]{#2}\begingroup\raggedright\endgroup



\end{document}